\documentclass[a4paper,fleqn]{cas-dc}

\usepackage[authoryear,longnamesfirst]{natbib}
\usepackage{times}
\usepackage{algorithm}
\usepackage{algorithmic}
\usepackage{bm}
\usepackage[switch]{lineno}
\usepackage{tabularx}
\usepackage{microtype}
\usepackage{subfigure}
\usepackage{longtable,ragged2e}
\usepackage{caption}
\usepackage{changepage}
\usepackage{listings}
\usepackage{enumitem}

\lstdefinestyle{json}{
    language=JavaScript,                 
    morekeywords={true,false,null},     
    keywordstyle=\color{blue!90!black}\bfseries,
    stringstyle=\color{orange!90!black},
    commentstyle=\color{gray},
    basicstyle=\ttfamily\small,
    columns=flexible,
    keepspaces=true,
    showstringspaces=false,
    backgroundcolor=\color{gray!5},    
    frame=tb,                            
    rulecolor=\color{gray!50},
}

\usepackage{balance}
\graphicspath{{./fig/}}
\def\tsc#1{\csdef{#1}{\textsc{\lowercase{#1}}\xspace}}
\tsc{WGM}
\tsc{QE}

\begin{document}
\let\WriteBookmarks\relax
\def\floatpagepagefraction{1}
\def\textpagefraction{.001}

    \shorttitle{Hierarchical Control Framework Integrating LLMs with RL for Decarbonized HVAC Operation }    

\shortauthors{Dianyu Zhong et al.}

\title[mode=title]{Hierarchical Control Framework Integrating Large Language Models with Reinforcement Learning for Decarbonized HVAC Operation}



%



\author[1]{Dianyu Zhong}[type=editor,style=chinese,orcid=0000-0002-3262-4905]
\fnmark[1]
\ead{zhongdy15@tsinghua.org.cn}



\author[1]{Tian Xing}[type=editor,style=chinese]
 \fnmark[1]

\author[2,3]{Kailai Sun}[type=editor,style=chinese,orcid=0000-0003-1648-3409]


\cormark[1]

\ead{skl24@mit.edu}

\author[1]{Xu Yang}[type=editor,style=chinese, orcid=]

\author[2,3]{Heye Huang}[type=editor,style=chinese, orcid=]



\author[1]{Irfan Qaisar}[type=editor,style=chinese, orcid=0000-0002-4831-977X]

\author[4]{Tinggang Jia}[type=editor,style=chinese]

\author[4]{Shaobo Wang}[type=editor,style=chinese]





\author[1]{Qianchuan Zhao}[type=editor,style=chinese,orcid=0000-0002-7952-5621]

\cormark[1]

\ead{zhaoqc@mail.tsinghua.edu.cn}




\affiliation[1]{organization={Center for Intelligent and Networked Systems, Department of Automation, BNRist, Tsinghua University},
            city={Beijing},
            postcode={100084}, 
            country={China}}


\affiliation[2]{organization={Singapore-MIT Alliance for Research and Technology Centre (SMART), Massachusetts Institute of Technology},
            postcode={138602}, 
            country={Singapore}}
\cortext[1]{Corresponding authors.}

\affiliation[3]{organization={Urban Mobility Lab, Massachusetts Institute of Technology},
            addressline={Cambridge}, 
            state={MA 02139},
            country={United States}}


\affiliation[4]{organization={ Shanghai Electric Automation Group Co.,Ltd.},
            postcode={Mengzi Road, Huangpu District, Shanghai,  200023},
            country={China}}


\fntext[1]{These authors contributed equally to this work.}

\begin{abstract}
Heating, ventilation, and air conditioning (HVAC) systems account for a substantial share of building energy consumption. Environmental uncertainty and dynamic occupancy behavior bring challenges in decarbonized HVAC control.
Reinforcement learning (RL) can optimize long-horizon comfort--energy trade-offs but suffers from exponential action-space growth and inefficient exploration in multi-zone buildings. 
Large language models (LLMs) can encode semantic context and operational knowledge, yet when used alone they lack reliable closed-loop numerical optimization and may result in less reliable comfort--energy trade-offs.
To address these limitations, we propose a hierarchical control framework in which a fine-tuned LLM, trained on historical building operation data, generates state-dependent feasible action masks that prune the combinatorial joint action space into operationally plausible subsets.
A masked value-based RL agent then performs constrained optimization within this reduced space, improving exploration efficiency and training stability.
Evaluated in a high-fidelity simulator calibrated with real-world sensor and occupancy data from a 7-zone office building, the proposed method achieves a mean PPD of 7.30\%, corresponding to reductions of 39.1\% relative to DQN, the best vanilla RL baseline in comfort, and 53.1\% relative to the best vanilla LLM baseline, while reducing daily HVAC energy use to 140.90~kWh, lower than all vanilla RL baselines. The results suggest that LLM-guided action masking is a promising pathway toward efficient multi-zone HVAC control.
\end{abstract}

\maketitle
\section{Introduction}
\label{section1}

Buildings are pivotal in the global transition towards a sustainable future. According to the recent Global Status Report for Buildings and Construction 2024/2025 ~\cite{2024un}, the building sector remains a key driver of the climate crisis, consuming 32\% of global energy and contributing 34\% of global CO2 emissions. Furthermore, the Global Buildings Climate Tracker indicates that the sector is still not aligned with the decarbonization pathway toward 2050, underscoring that advancing building energy efficiency remains a global priority for effectively mitigating the energy and climate crises \cite{sun2020review,ma2023energy}. Beyond energy and carbon concerns, the U.S. Environmental Protection Agency has noted that people spend nearly 90\% of their lives indoors \cite{EPA}. Consequently, the quality of indoor environments directly affects health, productivity, and comfort across multiple dimensions—thermal, visual, air quality, and acoustics. Among these, thermal comfort (defined by ASHRAE Standard 55 \cite{ashrae2023standard55}) plays a crucial role in overall indoor environmental quality, as it bridges physiological needs with subjective satisfaction.

To achieve the Net Zero target by 2050, incorporating intelligent and efficient control in building energy management systems\cite{wei2025optimal} is essential for reducing carbon emissions and improving human comfort \cite{SOLEIMANIJAVID2024113958}. However, challenges like varying sources of uncertainty, such as climate and weather conditions, and dynamic occupancy behavior, can significantly influence building energy consumption patterns. To address this, IEA-EBC Annex 66 \cite{IEA66} highlights occupant behaviors such as thermostat use, window opening and lighting affect building energy performance, and provides methods to model them. Annex 79 \cite{annex79} advances Annex 66 by applying sensing and machine learning to integrate occupant behavior into design and operation for better efficiency and comfort. Thus, occupant-centric control (OCC) has been developed as a transformative strategy that integrates real-time sensing of indoor environmental quality, occupancy information \cite{SUN2022111593}, and occupant-building interactions to dynamically adjust building operations, such as HVAC systems ~\cite{nagy2023ten,xing2022honeycomb}. 

A broad range of control strategies have been explored for HVAC and building energy management, including rule-based control, model predictive control (MPC) \cite{ajagekar2025decarbonization}, and reinforcement learning (RL). 

Among these, RL is attractive because it can optimize long-horizon comfort--energy trade-offs directly from interaction data. However, deploying RL in real multi-zone HVAC systems faces two major challenges. First, the joint action space grows exponentially with the number of zones and discrete actuation levels, leading to severe sample inefficiency and unstable learning in high-dimensional environments. Second, RL relies on trial-and-error exploration. While such exploration may be acceptable in simulation, comfort-violating or operationally aggressive actions are undesirable in occupied buildings because they may cause occupant discomfort and unnecessary equipment cycling. These factors have limited the practical adoption of RL-based HVAC controllers.

Recently, large language models (LLMs) have emerged as a promising tool for building management because they can interpret contextual information, encode human operational knowledge, and reason over complex semantic constraints~\cite{chang2024survey,sun2025review,jiang2024eplus,zhang2024large}. 
Their robustness and generalization capabilities make them attractive for real-time decision support in building systems, especially under partial observability and non-stationary occupancy patterns.
However, using LLMs as stand-alone HVAC controllers remains problematic.

Although LLMs can generate plausible control suggestions, they do not provide reliable closed-loop numerical optimization under coupled thermal dynamics and delayed rewards. 
As a result, direct LLM control may result in less reliable comfort–energy trade-offs, such as conservative over-conditioning or insufficient conditioning, and it lacks a principled mechanism for improvement through reward feedback. 

These complementary limitations motivate a hierarchical control paradigm. In particular, multi-zone HVAC control requires a framework that can leverage historical operational knowledge to constrain RL exploration without replacing reward-driven optimization. 
In this work, we propose a hierarchical LLM--RL control framework for multi-zone HVAC systems. 
An LLM fine-tuned on historical building operation data generates state-dependent feasible action masks that prune the combinatorial joint action space into historically grounded, operationally plausible subsets. 
A masked value-based RL agent then performs constrained optimization within this reduced space, improving training efficiency and stability. 
The proposed framework is trained and evaluated in a high-fidelity HVAC simulator calibrated using real-world sensor and occupancy data from a 7-zone office building.


The main contributions of this work are as follows:

\begin{itemize}
\item We propose a hierarchical LLM-RL control framework for multi-zone HVAC operation, in which a fine-tuned LLM generates state-dependent feasible action masks and a masked RL agent performs constrained optimization within the reduced action space.
\item We finetune a general LLM with domain-specific real-world historical operation data in buildings, capturing effective HVAC control patterns and guiding a downstream RL agent for efficient learning.
\item We develop and calibrate a high-fidelity HVAC simulator using real-world sensor and occupancy data from a 7-zone office building, providing a realistic testbed for controller training and comfort-energy evaluation.
\item Extensive experiments in the calibrated 7-zone case study show that the proposed framework substantially reduces the effective action space and improves RL exploration efficiency and training stability. It achieves a mean PPD of 7.30\%, corresponding to reductions of 39.1\% relative to DQN and 53.1\% relative to the best vanilla LLM baseline, while reducing daily HVAC energy use to 140.90~kWh, lower than all vanilla RL baselines.

\end{itemize}

 \begin{table}[h]
\centering
\begin{tabular}{|ll|}
\hline 
\multicolumn{2}{|l|}{\textbf{List of Abbreviations}}  \\
& \\
GenAI & Generative Artificial Intelligence\\
LLMs & Large Language Models \\
RL & Reinforcement Learning \\
A2C & Advantage Actor-Critic \\
BEMS & Building Energy Management Systems \\
CO2 & Carbon Dioxide \\
CPU & Central Processing Unit \\
DQN & Deep Q-Network \\
DRL & Deep Reinforcement Learning \\
FCU & Fan Coil Unit \\
GPU & Graphics Processing Unit \\
HVAC & Heating, Ventilation, and Air Conditioning \\
MLP & Multilayer Perceptron \\
OCC & Occupant-Centric Control \\
OTTV & Overall Thermal Transfer Value \\ 
PIR & Passive Infrared sensor \\
PMV & Predicted Mean Vote \\
PPD & Predicted Percentage of Dissatisfied \\
PPO & Proximal Policy Optimization \\
\hline
\end{tabular}
\end{table}

The remainder of this paper is organized as follows. Section~\ref{sectionL} reviews related work. Section~\ref{sec:method} presents the proposed hierarchical LLM--RL framework. Section~\ref{sec:exp setting} describes the experimental setup, Section~\ref{sec:results} reports the results and discussion, Section~\ref{sec:limitations} discusses limitations, and Section~\ref{sec:conclusion} concludes the paper.

\section{Literature Review}
\label{sectionL}

\subsection{Occupant-centric HVAC control}

Reliable occupancy information is a prerequisite for occupant-centric building operation because it directly influences both energy management and thermal comfort assessment \cite{sun2022fusion,xing2022honeycomb,PANG2020115727}. Existing studies on occupancy-related information can be broadly organized into three levels: presence detection, occupant counting, and activity or behavior inference. Presence detection, which determines whether a space is occupied, remains the most widely used level in practice. Passive Infrared (PIR) sensing is commonly adopted for this purpose in buildings \cite{FATEHIKARJOU2024114852}, and its performance can be improved by integrating Wireless Sensing Network modules or Radio Frequency Identification technologies \cite{SOLEIMANIJAVID2024113958}. Research has subsequently extended from binary presence recognition to occupant counting. Carbon-dioxide-based methods are frequently used for counting, although their accuracy can be affected by sensor placement as well as by window and HVAC operation \cite{hobson2019opportunistic,jung2019human}. To improve robustness, recent work has explored vision-based sensing \cite{sun2022fusion} together with other data sources, such as Global Positioning System information and WiFi signals \cite{zhu2024wisa}. Compared with presence detection and counting, activity-level understanding is still less mature, although motion sensing \cite{sun2020review} and vision-based methods \cite{tien2020vision} have enabled more detailed characterization of occupant behaviors.

From the control perspective, most occupant-centric strategies still rely primarily on presence detection \cite{HOBSON2025115087,QAISAR2025112322, brooks2015experimental, liu2023advanced} or occupant counting \cite{CELINEJACOB2024114497, hobson2019opportunistic, jung2019human}, whereas richer behavioral descriptors are much less frequently incorporated. In HVAC applications, presence information has commonly been used to implement temperature setpoint scheduling \cite{ye2021energy,pang2021much}. In broader building operation, the same binary occupancy logic has also supported vacancy-based lighting control \cite{saha2019occupancy,salimi2019critical}. Occupant counting obtained from sensors such as RGB cameras and CO$_2$ devices has further been introduced into predictive control of air-conditioning systems and outdoor air handling units \cite{wang2017predictive}. More recent studies have started to move beyond counting by using contextual and historical information to predict occupant activities and improve control performance \cite{turley2020development}. Activity data collected from wearable devices have also been used to reduce thermal discomfort by 10.9\% without increasing energy consumption \cite{jung2022occupant}. Overall, the literature shows a clear shift from coarse occupancy awareness toward richer behavioral information, which provides a stronger foundation for adaptive and occupant-centric HVAC control.

\subsection{Reinforcement learning applications in HVAC systems}

Reinforcement learning has attracted sustained attention in building control because it can optimize comfort and energy performance through sequential interaction with dynamic environments \cite{park2020hvaclearn}. A number of studies have demonstrated its applicability to HVAC operation in both simulated and real settings. For example, Sun et al.~\cite{sun2024individual} used deep reinforcement learning for residential air-conditioning automation and reported about 40\% energy savings in winter. Silvestri et al.~\cite{silvestri2025practical} deployed a model-free deep reinforcement learning controller enhanced by imitation learning in a real building, achieving around 40\% energy savings together with up to 43\% fewer temperature violations than rule-based control. Zhang et al.~\cite{zhang2019whole} applied an Asynchronous Advantage Actor-Critic agent in an office building, while Chen et al.~\cite{chen2019gnu} embedded a differentiable model predictive control policy into a deep reinforcement learning framework and implemented it in a conference room for three weeks. Qiu et al.~\cite{qiu2022chilled} combined reinforcement learning with expert knowledge for chilled water temperature resetting in a real HVAC system. Liu et al.~\cite{liu2022multi} further proposed a multi-step prediction-oriented deep RL method that reduced power consumption by 12.79\% relative to conventional on/off control.

As the focus of building control has expanded from single variables to coupled multi-zone operation, RL has also been extended to higher-dimensional settings. Liu et al.~\cite{liu2024enhancing} developed a multi-agent deep reinforcement learning approach for occupant-centric multi-zone HVAC control and reported a 51.09\% reduction in electricity cost compared with rule-based control while maintaining thermal comfort. Nguyen et al.~\cite{nguyen2024modelling} proposed a Phasic Policy Gradient framework for multi-zone HVAC control, yielding 2--14\% energy savings, improved temperature comfort, and faster convergence than conventional methods. In related multivariate occupant-centric building control, Park et al.~\cite{park2019lightlearn} applied RL to lighting control in five office spaces and showed that the learned controller could adapt to individual behaviors and indoor environmental conditions to determine personalized setpoints. Lei et al.~\cite{lei2022practical} also presented a practical deep reinforcement learning framework for multivariate occupant-centric control that jointly addressed occupant presence and personalized thermal comfort. Taken together, these studies confirm the promise of RL for building control, while also indicating that control complexity grows rapidly as more zones, variables, and occupancy-driven interactions are considered.

\subsection{Large language models in building energy applications}
Recent studies indicate that large language models (LLMs) are being rapidly integrated into building energy applications, with prompt engineering, fine-tuning, inference-based reasoning, and agent workflows emerging as major methodological paradigms~\cite{arslan2025large,liu2025large, chang2024survey}. 
Beyond their role as natural-language interfaces, LLMs are increasingly being explored as carriers of semantic and operational knowledge for building-related decision support\cite{veloso2025forming, burgueno2024human, lin2024bitsa, jin2024democratizing, jurivsevic2024exploring, zhang2024generative, zhang2024large, ashayeri2024unraveling, sadick2025did}.
In this sense, LLMs provide a new pathway for knowledge-guided building control: instead of relying solely on handcrafted rules or purely numerical policies, controllers can leverage contextual semantics, domain knowledge, and historical operational patterns to improve decision quality and adaptability.

Within the HVAC and thermal comfort domain, existing studies have explored LLMs and related generative-AI techniques for real-time control, predictive control, scalable supervisory operation, fault detection and diagnostics, and personalized comfort services. Several studies have shown that LLM-based or GenAI-enhanced methods can improve both energy efficiency and occupant comfort in HVAC operation~\cite{ahn2023alternative38,zhu2025heating45,sawada2024office48,sawada2025office49}, while others have proposed scalable and predictive control schemes to improve deployment potential in larger or more complex systems~\cite{ma2024beforegan39,zhu2021fast44,li2025large50}. Related research has also investigated domain-adapted LLMs and time-series modeling for fault detection and diagnostics~\cite{zhang2025domain41,liu2025integrating46}, transfer learning and benchmarking for model generalizability~\cite{kadamala2024enhancing42}, and multi-agent or GenAI-based frameworks for personalized thermal comfort and climate control~\cite{arslan2024decision40,liu2025integrating46}. Collectively, these studies suggest that LLMs are becoming a promising component of intelligent building operation, especially when control decisions depend on heterogeneous contextual information rather than only instantaneous numerical states.

Despite this rapid progress, an important gap remains in the literature. Many existing studies use LLMs either as stand-alone decision generators or as high-level assistants, whereas fewer works explicitly integrate LLM-derived knowledge with closed-loop optimization for sequential HVAC control. Moreover, although fine-tuning and domain adaptation are increasingly adopted, the exploration of historical building-operation data to inject building-specific knowledge into downstream control remains relatively underexplored. This limitation is particularly important in multi-zone HVAC systems, where the combinatorial action space makes unguided exploration inefficient and operationally implausible. These observations motivate a hybrid knowledge-data-driven control paradigm in which LLMs provide historically grounded and meaningful feasible-action knowledge, while reinforcement learning remains responsible for long-horizon numerical optimization under coupled thermal dynamics and delayed rewards. Against this background, this study treats the LLM not as a direct controller, but as a generator of state-dependent feasible action masks that guide downstream RL optimization.

\section{Methodology}
\label{sec:method}

\begin{figure*}[t]
    \centerline{
    \includegraphics[width=2\columnwidth, trim=0cm 0cm 0cm 0cm, clip]{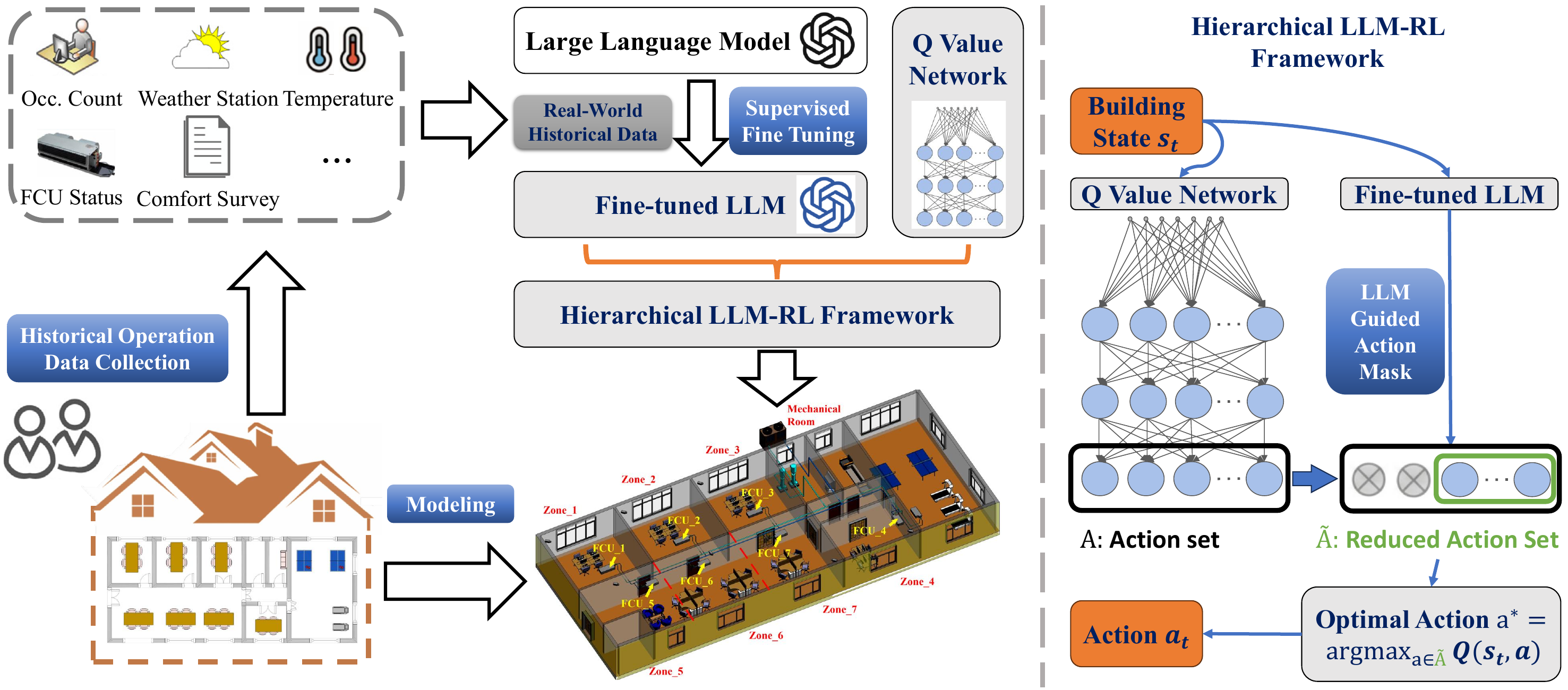}
    }
    \caption{Overview of the proposed hierarchical LLM--RL framework for multi-zone HVAC control. Historical operation data are used to construct feasible-action labels and supervise the LLM, while the calibrated simulator provides the training and evaluation environment for the RL agent. During online control, the fine-tuned LLM generates a state-dependent feasible action mask, and the masked DQN selects the final joint FCU action within the reduced action space.}
    \label{fig:framework}
\end{figure*}

This section describes the proposed {hierarchical Large Language Model--Reinforcement Learning (LLM-RL) control framework} for HVAC control in buildings, designed to jointly optimize thermal comfort and energy efficiency. 
As illustrated in Fig.~\ref{fig:framework}, the framework consists of two decision-making layers: an upper LLM layer and a lower RL layer. The upper layer employs a fine-tuned LLM, trained on real-world historical operation data, to generate state-dependent feasible action masks that characterize human-like and operationally admissible control patterns. The lower layer is a value-based RL agent that searches for the optimal action only within the feasible action region specified by the LLM. 

Accordingly, the hierarchical decision rule can be written as
\begin{equation}
\left\{
\begin{aligned}
\widetilde{\mathcal{A}}(s_t) &= 
\underbrace{\Psi(x_t)}_{\text{LLM-based action mask generation}},\\
a_t &= 
\underbrace{\arg\max_{a\in \widetilde{\mathcal{A}}(s_t)} Q(s_t,a;\theta)}_{\text{RL action selection}}.
\end{aligned}
\right.
\end{equation}

where $\Psi(\cdot)$ denotes the fine-tuned LLM, $x_t$ is the structured prompt constructed from the current and recent observations, and $\widetilde{\mathcal{A}}(s_t)$ is the LLM-generated feasible joint action set at time step $t$. In addition, $Q(s_t,a;\theta)$ denotes the action-value function learned by the RL agent with parameters $\theta$, and $a_t$ is the action selected by maximizing the Q-value over the feasible action set. In this way, the LLM defines the admissible decision boundary through action masking, while the RL agent performs optimal control selection within that reduced space.

The framework operates in two stages. 
In the offline stage, real-world historical building operation data are used to construct feasible-action labels. 
These labels are used to fine-tune a pre-trained LLM so that it can map structured building-state prompts to zone-wise feasible fan-speed sets.
In the online stage, recent observations are serialized into a structured prompt and fed to the fine-tuned LLM to generate feasible action masks. The RL agent then selects the final control action by maximizing the Q-value over the masked action space, while the calibrated simulator provides state transitions and rewards for closed-loop learning and evaluation.

The proposed hierarchical LLM–RL control framework comprises four major components, detailed in the subsequent sections:

\begin{itemize}
\item Control problem formulation (Section~\ref{sec:problem_formulation}).
\item Offline data collection and LLM fine-tuning (Sections~\ref{sec:data} and~\ref{sec:sft}).
\item Online control with LLM-guided action masking and masked RL decision-making (Section~\ref{sec:masked_dqn}).
\item Building HVAC simulator construction for training and evaluation (Section~\ref{sec:simulator}).
\end{itemize}

The following subsections describe each component in detail.

\subsection{Control Problem Formulation}
\label{sec:problem_formulation}

We formulate multi-zone HVAC operation as a finite-horizon sequential decision problem with state-dependent action constraints. At each control step $t$ with interval $\Delta t=5$ minutes, the controller observes the current building state $s_t \in \mathcal{S}$ and selects a joint control action $a_t$ for all controllable fan coil units (FCUs). The transition from $s_t$ to $s_{t+1}$ is governed by the coupled building thermal and HVAC dynamics described in Section~\ref{sec:simulator}.

The state $s_t$ summarizes the information available from the sensing and management system at time $t$, including zone indoor temperatures, zone-level occupancy counts, outdoor thermal conditions, recent FCU operating states, and auxiliary HVAC and temporal features. For notation convenience, we write
\begin{equation}
s_t=\bigl[\mathbf{T}^{\mathrm{zone}}_t,\ \mathbf{n}_t,\ T^{\mathrm{out}}_t,\ \mathbf{a}_{t-1},\ \mathbf{z}_t\bigr],
\end{equation}
where $\mathbf{T}^{\mathrm{zone}}_t=[T_{t,1},\dots,T_{t,J}]$ denotes the vector of indoor temperatures for the $J$ zones ($J=7$ in the present case study), $\mathbf{n}_t=[n_{t,1},\dots,n_{t,J}]$ denotes the corresponding vector of zone-level occupancy counts, and $\mathbf{a}_{t-1}=[a_{t-1,1},\dots,a_{t-1,J}]$ denotes the previous FCU fan-speed settings. $T^{\mathrm{out}}_t$ denotes the outdoor temperature, and $\mathbf{z}_t$ collects auxiliary HVAC measurements and time/context features.

In the proposed framework, recent observations are further serialized over a short temporal window for the LLM module, whereas the RL module uses the current decision state together with the LLM-generated action mask for value estimation and action selection.

For a building with $J$ controllable FCUs, the control action at time $t$ is the joint discrete fan-speed vector
\begin{equation}
a_t=[a_{t,1},a_{t,2},\dots,a_{t,J}], \qquad a_{t,j}\in \mathcal{L},
\end{equation}
where $\mathcal{L}=\{0,1,2,3\}$ denotes the available fan-speed levels, with $0$ representing the off state and $1$--$3$ representing increasing fan speeds. The full joint action space is therefore
\begin{equation}
\mathcal{A}=\mathcal{L}^{J},
\end{equation}
whose cardinality is $|\mathcal{A}|=4^7=16384$ for the studied 7-zone building.

Unlike unconstrained RL, the proposed framework operates over a state-dependent feasible action subset generated by the fine-tuned LLM. Given the current information at time $t$, the LLM outputs a feasible fan-speed set $M_j(s_t)\subseteq \mathcal{L}$ for each FCU $j$. These zone-wise feasible sets induce the valid joint action subset
\begin{equation}
\widetilde{\mathcal{A}}(s_t)=\prod_{j=1}^{J} M_j(s_t)\subseteq \mathcal{A},
\end{equation}
which constrains downstream exploration and action selection.

The control objective is to maximize the expected discounted return while balancing thermal comfort and HVAC energy use:
\begin{equation}
\max_{\pi}\ \mathbb{E}_{\pi}\!\left[\sum_{t=0}^{T-1}\gamma^t r_t\right]
\qquad \text{s.t.} \qquad
a_t\in \widetilde{\mathcal{A}}(s_t),\ \forall t.
\end{equation}
Here, the per-step reward $r_t$ penalizes occupancy-weighted thermal discomfort during occupied periods together with HVAC energy use, while the detailed comfort and energy definitions are provided in Section~\ref{sec:metrics}. Under this formulation, the LLM is responsible for generating state-dependent feasible action masks, and the RL agent performs policy optimization within the masked action space.

\subsection{Real-World Historical Data Collection}
\label{sec:data}

To support both LLM training and simulator calibration, historical operational data were collected from a commercial office building in Hebei Province, China, from August 7 to August 22, 2021~\cite{xing2022honeycomb}. Measurements were logged at 1-minute intervals during operational hours (9:00--19:00), while FCU control actions were executed every 5 minutes. Over 9{,}000 synchronized timestamps were recorded, covering thermal-zone conditions, occupancy states, and HVAC operating variables.

The collected dataset includes outdoor temperature, zone air temperatures, FCU fan states, supply and return water temperatures, supply and return pressures, and zone-level occupancy counts. These variables provide the empirical basis for the state representation in Section~\ref{sec:problem_formulation}, for the kNN-based feasible-action labeling described in Section~\ref{sec:sft}, and for simulator calibration in Section~\ref{sec:simulator}. The key variables are summarized in Table~\ref{tab:room-state-keys}.

\begin{table}[h]
    \centering
    \caption{Historical Data Collection.}
    \label{tab:room-state-keys}
    \begin{tabular}{ll}
        \toprule
        \textbf{Variable} & \textbf{Description} \\
        \midrule
        {outdoor\_temp} & Outdoor air temp ($^\circ$C) \\
        \midrule
         For $i=1,\dots,7$ & \\
         \quad {zone\_temp\_i} &  Zone $i$ air temp ($^\circ$C)\\
        \quad FCU\_fan\_i &  $\text{FCU}_i$ fan speed mode \\
        \quad {supply\_temp\_i} &  $\text{FCU}_i$ supply temp ($^\circ$C) \\
        \quad {return\_temp\_i} &  $\text{FCU}_i$ return temp ($^\circ$C) \\
        \quad {supply\_pressure\_i} & $\text{FCU}_i$ supply pressure (kPa) \\
        \quad {return\_pressure\_i}&   $\text{FCU}_i$ return pressure (kPa) \\
        \quad {occupant\_num\_i} &  Zone $i$ Occ. count \\
        \bottomrule
    \end{tabular}
\end{table}

During the data-collection period, the baseline control strategy combined occupancy-based automation with manual overrides. In automatic mode, FCUs were activated or deactivated according to real-time occupancy detection, while local thermostats or the building management system could temporarily override the automated control. As a result, the logged trajectories reflect historically observed operation patterns under real building use rather than synthetic expert demonstrations. Detailed case-study descriptions and evaluation settings are provided in Section~\ref{subsec:case_study} and Section~\ref{sec:metrics}, respectively.

\subsection{Supervised Fine-Tuning of the LLM}
\label{sec:sft}

To enable the LLM to generate state-dependent feasible action masks for the downstream controller, we perform supervised fine-tuning (SFT) on the historical HVAC operational data. This stage consists of two steps: constructing feasible-action labels from neighborhoods of similar states and adapting a pre-trained LLaMA model via Low-Rank Adaptation (LoRA) to predict these labels from structured prompts.

\subsubsection{Construction of SFT Dataset via kNN-Derived Feasible Action Sets}

The raw historical dataset contains state--action pairs $\{(s_i,a_i)\}_{i=1}^N$, where $s_i$ denotes the building state at time step $i$ and $a_i$ is the executed joint FCU fan-speed command. In HVAC control, however, more than one action may be acceptable under similar comfort--energy conditions. To expose this multimodality to the LLM, we augment each logged action into a data-driven feasible action set derived from a neighborhood of similar historical states.

For each state $s_i$, we first define a weighted Euclidean distance in the state space:
\begin{equation}
d(s_i,s_l)=\sqrt{\sum_{q} w_q\bigl(s_i^{(q)}-s_l^{(q)}\bigr)^2},
\end{equation}
where $s_i^{(q)}$ denotes the $q$-th feature of state $s_i$ and $w_q$ is the corresponding feature weight. The neighborhood of $s_i$ is then defined as
\begin{equation}
\mathcal{N}(i)=\operatorname{kNN}_{k}(s_i),
\end{equation}
which contains the indices of the $k$ most similar historical states.

Given the fan-speed level set $\mathcal{L}=\{0,1,2,3\}$, we compute, for each FCU $j\in\{1,\dots,J\}$ and level $l\in\mathcal{L}$, the empirical frequency of applying level $l$ within the neighborhood of $s_i$:
\begin{equation}
f_{j,l}^{(i)}=
\frac{1}{|\mathcal{N}(i)|}
\sum_{m\in\mathcal{N}(i)} \mathbb{1}\bigl(a_{m,j}=l\bigr),
\end{equation}
where $a_{m,j}$ is the fan-speed level of FCU $j$ in the historical action $a_m$.

We then define the feasible fan-speed set for FCU $j$ at state $s_i$ by thresholding these empirical frequencies:
\begin{equation}
M_j(s_i)=\bigl\{\, l\in\mathcal{L}\ \big|\ f_{j,l}^{(i)}\ge \tau \,\bigr\},
\end{equation}
where $\tau\in[0,1]$ is a small frequency threshold used to exclude extremely rare actions.
In all experiments, we use a fixed neighborhood size $k = 50$ and a frequency threshold $\tau = 0.05$ when constructing feasible action sets.
The corresponding binary labels are
\begin{equation}
y_{i,j,l}=
\begin{cases}
1, & l\in M_j(s_i),\\
0, & \text{otherwise}.
\end{cases}
\end{equation}

Following the prompting scheme used at inference time, each SFT sample consists of two parts: an input prompt $x_i$ and a target output $o_i$. The input prompt serializes the recent building-state sequence from $t-4$ to $t$, including temperatures, occupancies, previous fan actions, and time/context information. 
The target output is a JSON object with two fields: \texttt{analysis}, generated from a fixed rule-based template derived from the observation window and the feasible-action statistics, and \texttt{recommendations}, which encodes the zone-wise feasible action sets.
During online control, only the \texttt{recommendations} field is used by the RL controller.
Collecting all such pairs yields the SFT dataset
\begin{equation}
\mathcal{D}_{\mathrm{SFT}}=\bigl\{(x_i,o_i)\bigr\}_{i=1}^{N}.
\end{equation}

\subsubsection{LoRA-Based Fine-Tuning of LLaMA for Mask Prediction}

Given $\mathcal{D}_{\mathrm{SFT}}$, we fine-tune a pre-trained LLaMA model to approximate the mapping from structured prompt to zone-wise feasible action sets. To preserve the general language capabilities of the base model while keeping the adaptation parameter-efficient, we adopt LoRA.

For a weight matrix $W_0\in\mathbb{R}^{d_{\mathrm{out}}\times d_{\mathrm{in}}}$ in the base model, LoRA re-parameterizes the adapted weight as
\begin{equation}
W=W_0+\Delta W,
\qquad
\Delta W=BA,
\end{equation}
where $A\in\mathbb{R}^{r\times d_{\mathrm{in}}}$ and $B\in\mathbb{R}^{d_{\mathrm{out}}\times r}$ are trainable low-rank matrices with rank $r\ll \min(d_{\mathrm{in}},d_{\mathrm{out}})$. During SFT, only the LoRA parameters are updated, whereas the original model weights remain frozen.

We perform standard teacher-forced next-token prediction. Each training pair $(x_i,o_i)$ is concatenated into one sequence, and the SFT objective minimizes the negative log-likelihood of the target tokens:
\begin{equation}
\mathcal{L}_{\mathrm{SFT}}(\phi)=
-\frac{1}{N}\sum_{i=1}^{N}\log p_{\phi}\bigl(o_i\,|\,x_i\bigr),
\end{equation}
where $\phi$ denotes the LoRA parameters. In practice, the loss is computed over the target portion of the sequence, namely the generated analysis text and the structured \texttt{recommendations} field. Detailed training configurations are provided in Section~\ref{subsubsec:q3_sft_curve} and Table~\ref{tab:sft_core_config}.

At inference time, the fine-tuned model receives the same prompt structure but without ground-truth labels and autoregressively generates a JSON object. The \texttt{recommendations} field is parsed to recover the zone-wise feasible sets $\{M_j(s_t)\}_{j=1}^{J}$, which in turn induce the masked action space $\widetilde{\mathcal{A}}(s_t)$ used by the RL controller.

\subsection{Online Control with LLM-guided Action Masking}

After offline SFT, the LLM is embedded in the online control loop to provide state-dependent action masks for the RL agent. This section describes how numerical building states are converted into structured LLM inputs and how the resulting masks are incorporated into masked DQN decision-making.

\subsubsection{Structured Prompting and LLM-guided Action Mask}

\begin{figure}
    \centerline{
    \includegraphics[width=\columnwidth, trim=0cm 0.8cm 0cm 0cm, clip]{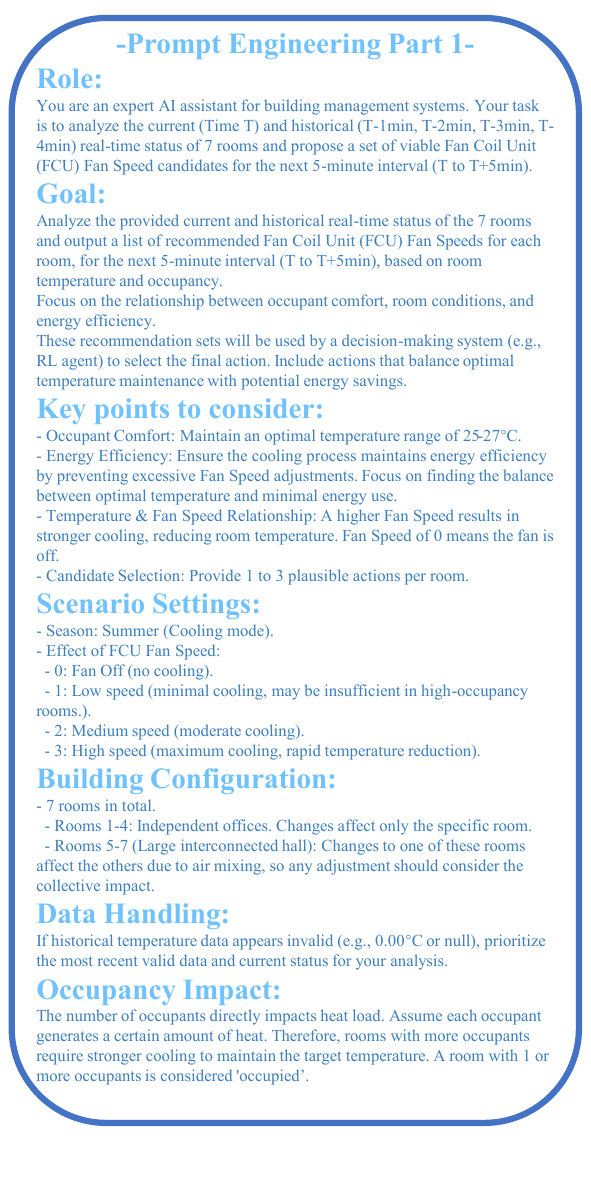}
    }
    \caption{System instruction component of the prompt. This part specifies the control objective, operational context, and domain knowledge on HVAC actuation and spatial topology.}
    \label{fig: prompt design part 1}
\end{figure}

\begin{figure}
    \centerline{
    \includegraphics[width=\columnwidth, trim=0cm 0cm 0cm 0cm, clip]{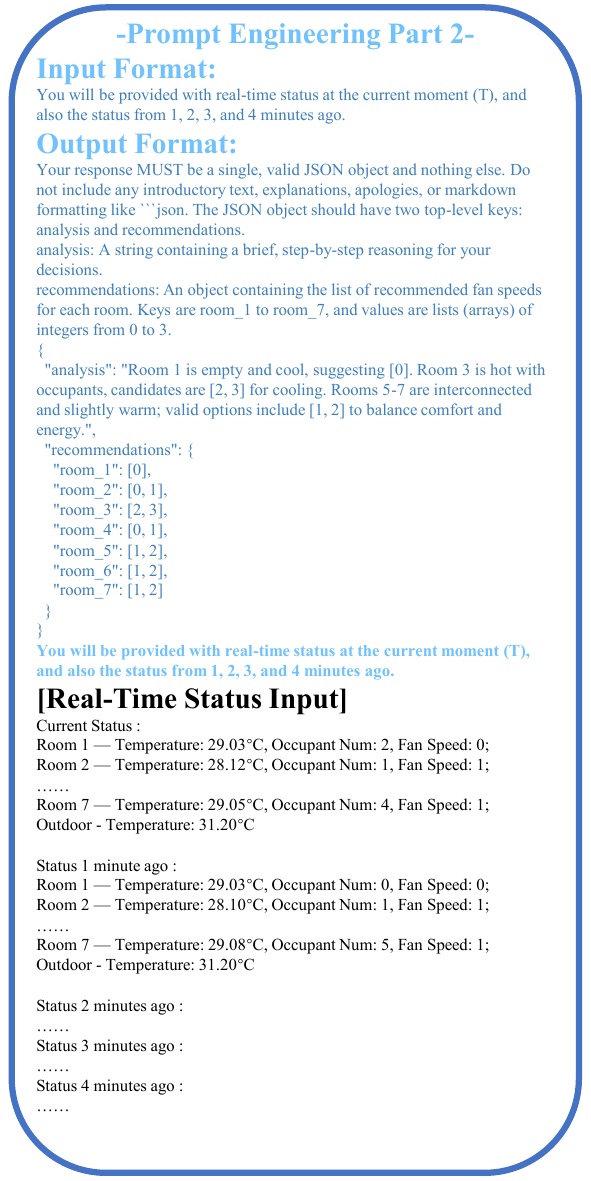}
    }
    \caption{Formatting and state-serialization component of the prompt. The prompt enforces a JSON output schema and represents recent observations over a short temporal window, enabling the LLM to produce structured feasible fan-speed recommendations.}
    \label{fig: prompt design part 2}
\end{figure}

To interface numerical building states with the language model, we design a structured prompting mechanism that encodes control objectives, environmental context, and recent temporal dynamics. As illustrated in Fig.~\ref{fig: prompt design part 1} and Fig.~\ref{fig: prompt design part 2}, the prompt is constructed to translate recent building observations into a machine-readable instruction that supports structured feasible-action generation.

The prompt has four key elements. First, the system instruction defines the LLM as a building-management assistant and states the dual control objective of maintaining thermal comfort while avoiding unnecessary HVAC energy use. Second, the prompt embeds building-specific domain knowledge, including the mapping between fan-speed levels and cooling intensity, the distinction between relatively independent zones (Rooms 1--4) and thermally coupled zones (Rooms 5--7), and the role of occupancy in shaping cooling demand. Third, the input state is serialized over the most recent five control steps, from $t-4$ to $t$, so that the model can infer short-term trends rather than relying only on the current snapshot. Fourth, the prompt enforces a strict JSON output format to facilitate downstream parsing.

At time step $t$, the resulting prompt $x_t$ contains recent zone temperatures, occupancy counts, previous fan actions, outdoor conditions, and time/context features. The fine-tuned LLM returns a JSON object with two fields: a short textual analysis and a \texttt{recommendations} field. Only the \texttt{recommendations} field is used by the controller. It specifies, for each FCU $j$, a subset $M_j(s_t)\subseteq \mathcal{L}$ of feasible fan-speed levels. These zone-wise subsets are then converted into the joint feasible action space
\begin{equation}
\widetilde{\mathcal{A}}(s_t)=\prod_{j=1}^{J} M_j(s_t),
\end{equation}
which induces the LLM-guided binary action mask used for online control.

\subsubsection{Deep Reinforcement Learning under Action Masking}
\label{sec:masked_dqn}

In the studied multi-zone HVAC system, the joint action space grows exponentially with the number of controllable zones and discrete fan-speed levels. Standard value-based RL therefore faces inefficient exploration in the full combinatorial action space. In this work, the downstream RL controller is implemented as a DQN agent whose decision process is constrained by the LLM-generated feasible action mask.

For each joint action $a\in\mathcal{A}$, we define a binary validity indicator
\begin{equation}
m_a(s_t)=
\begin{cases}
1, & a\in \widetilde{\mathcal{A}}(s_t),\\
0, & a\notin \widetilde{\mathcal{A}}(s_t),
\end{cases}
\end{equation}
and collect these indicators into a binary mask vector $\mathbf{m}(s_t)\in\{0,1\}^{|\mathcal{A}|}$. 

The control action is then selected according to the constrained greedy policy
\begin{equation}
a_t=\arg\max_{a\in \widetilde{\mathcal{A}}(s_t)} Q(s_t,a;\theta),
\label{eq:masked_argmax}
\end{equation}
where $Q(s_t,a;\theta)$ is the Q-function parameterized by $\theta$.

In practice, invalid actions are suppressed by Q-value masking. Let $\mathbf{q}(s_t)\in\mathbb{R}^{|\mathcal{A}|}$ denote the raw Q-values produced by the network.
The masked Q-values are computed as
\begin{equation}
\mathbf{q}^{\mathrm{masked}}(s_t)
=
\mathbf{q}(s_t)
-
C\bigl(\mathbf{1}-\mathbf{m}(s_t)\bigr),
\end{equation}
where $C\gg 0$ is a large positive constant. Action selection, whether greedy or $\epsilon$-greedy, is then restricted to the feasible action subset.

The Bellman target is also computed under the next-state mask:
\begin{equation}
y_t=r_t+(1-d_t)\gamma \max_{a' \in \widetilde{\mathcal{A}}(s_{t+1})} Q(s_{t+1},a';\theta^-),
\end{equation}
where $d_t$ is the terminal indicator and $\theta^-$ denotes the target-network parameters. The DQN objective is
\begin{equation}
\mathcal{L}_{\mathrm{DQN}}(\theta)=
\mathbb{E}_{(s_t,a_t,r_t,s_{t+1},d_t)\sim \mathcal{D}}
\Bigl[
\bigl(y_t-Q(s_t,a_t;\theta)\bigr)^2
\Bigr],
\end{equation}
with $\mathcal{D}$ denoting the replay buffer.

The online decision loop is therefore as follows: the controller observes $s_t$, constructs the prompt $x_t$, queries the fine-tuned LLM to obtain $\widetilde{\mathcal{A}}(s_t)$, selects a valid action via masked DQN, executes the action in the simulator, and uses the resulting transition to update the value function. The corresponding training procedure is summarized in Algorithm~\ref{alg:llm_masked_dqn}.

\begin{algorithm}[t]
\caption{LLM-guided Masked DQN Training for Multi-zone HVAC Control}
\label{alg:llm_masked_dqn}
\begin{algorithmic}[1]
\REQUIRE Fine-tuned LLM $\Psi$, full action space $\mathcal{A}$, exploration schedule $\epsilon(t)$
\STATE Initialize online Q-network $Q(\cdot,\cdot;\theta)$ and target Q-network $Q(\cdot,\cdot;\theta^-)$
\STATE Initialize replay buffer $\mathcal{D}\leftarrow \emptyset$
\FOR{episode $k=1,2,\dots,K$}
    \STATE Reset simulator and observe initial state $s_0$
    \FOR{time step $t=0,1,\dots,T-1$}
        \STATE Construct structured prompt $x_t$ from recent observations
        \STATE Query LLM: $\{M_j(s_t)\}_{j=1}^{J}\leftarrow \Psi(x_t)$
        \STATE Form $\widetilde{\mathcal{A}}(s_t)=\prod_{j=1}^{J} M_j(s_t)$ and the binary mask $\mathbf{m}(s_t)$
        \STATE Select $a_t$ by $\epsilon$-greedy over $\widetilde{\mathcal{A}}(s_t)$ using masked Q-values
        \STATE Execute $a_t$ in the simulator and observe $r_t$, $s_{t+1}$, and $d_t$
        \STATE Store transition $(s_t,a_t,r_t,s_{t+1},d_t)$ in $\mathcal{D}$
        \IF{training update is triggered}
            \STATE Sample mini-batch $\mathcal{B}\sim \mathcal{D}$
            \FOR{each $(s_j,a_j,r_j,s_{j+1},d_j)\in \mathcal{B}$}
                \STATE Construct $x_{j+1}$ and query $\Psi(x_{j+1})$ to obtain $\widetilde{\mathcal{A}}(s_{j+1})$
                \STATE Compute masked target
                \[
                y_j=r_j+(1-d_j)\gamma \max_{a' \in \widetilde{\mathcal{A}}(s_{j+1})} Q(s_{j+1},a';\theta^-)
                \]
            \ENDFOR
            \STATE Update $\theta$ by minimizing the DQN loss over $\mathcal{B}$
            \STATE Periodically update $\theta^-$
        \ENDIF
    \ENDFOR
\ENDFOR
\STATE \textbf{Return} $\pi(s)=\arg\max_{a\in \widetilde{\mathcal{A}}(s)} Q(s,a;\theta)$
\end{algorithmic}
\end{algorithm}

\subsection{Building HVAC Simulator Construction}
\label{sec:simulator}

Based on the collected data and the designed system, we developed a high-fidelity simulator to replicate the building's thermal and energy dynamics, following the framework proposed by Yan et al.~\cite{yan2023protocol}. The simulator integrates real-world data to ensure alignment with the building's operational characteristics and consists of three main components: indoor zone models, HVAC equipment models, and a cooling water pipe network \cite{ZHONG2025116219,2025Coupling}.

The indoor zone model comprehensively accounts for multiple heat-gain components, including building envelope conduction, solar radiation, internal occupancy loads, inter-zone heat transfer, and the cooling/heating effect from terminal HVAC units. Heat transfer through the envelope is quantified using the Overall Thermal Transfer Value (OTTV) method \cite{KHANHPHUONG2024108616}. Occupant sensible heat release is modeled as the following function:
\begin{equation}
Q_{\text{occ},i} = \left( \frac{37 - T_{\text{in},i}}{37 - 24} \cdot q_{\text{p}} + q_{\text{d}} \right) \cdot n_{\text{p},i},
\end{equation}
where $q_{\text{p}}$ corresponds to the sensible heat emission of an average adult at 24°C, $q_{\text{d}}$ is an additional heat release correction, and $n_{\text{p},i}$ is the instantaneous number of occupants in the zone.

Inter-zone heat exchange between thermally connected zones is calculated as:
\begin{equation}
Q_{\text{int},i} = \sum_{j \in \mathcal{Z}_{\text{adj},i}} (T_{\text{in},j} - T_{\text{in},i}) \cdot \eta_{\text{adj},j},
\end{equation}
where $\mathcal{Z}_{\text{adj},i}$ is the set of zone \textit{i}'th adjacent zones, $T_{\text{in},j}$ is the air temperature of neighboring zone $j$, and $\eta_{\text{adj},j}$ represents the effective heat transfer coefficient of zone $j$.

The total thermal load of the zone is then obtained by aggregating all contributions:
\begin{equation}
Q_{\text{load},i} = Q_{\text{occ},i} + Q_{\text{int},i} + \sum_{w} OTTV_{i,w} \cdot A_{i,w},
\end{equation}
where $OTTV_{i,w}$ and $A_{i,w}$ are the OTTV value and the surface area of the $w$-th external wall of zone \textit{i}.

The sensible cooling or heating capacity delivered by supply air from fan coil units (FCUs) or other air-based terminals is expressed as:
\begin{equation}
\label{eq:q_supply}
Q_{\text{sup},i} = \rho_{\text{a},i} c_{p,\text{a}} \dot{V}_{\text{sup},i} (T_{\text{sup},i} - T_{\text{in},i}),
\end{equation}
where $\rho_{\text{a},i}$ and $c_{p,\text{a}}$ are the density and specific heat capacity of air, $ \dot{V}_{\text{sup},i}$ is the supply airflow rate of zone \textit{i}, and $T_{\text{sup},i}$ is the supply air temperature.

Finally, the instantaneous rate of change of zone air temperature is computed using a lumped capacitance approach:
\begin{equation}
\frac{dT_{\text{in},i}}{dt} = \frac{Q_{\text{load},i} + Q_{\text{sup},i}}{\rho_{\text{a},i} c_{p,\text{a}} V_{\text{zone},i}} \cdot \beta,
\end{equation}
where $V_{\text{zone},i}$ is the air volume of zone \textit{i} and $\beta \in [0.8, 1.2]$ is an empirical calibration factor that accounts for thermal stratification, furniture capacitance, and other non-ideal effects typically observed in real buildings.


The HVAC module comprises parameterized models for fan coil units (FCUs) and variable-speed circulating pumps. These components are responsible for terminal air distribution, coil heat transfer, system-side hydraulic performance, and overall electrical energy consumption. The fan power of FCUs under off-design airflow conditions is calculated using the fan similarity laws \cite{ashrae2021}:
\begin{equation}
    W_{\text{fcu},i}=(\frac{\dot{V}_{\text{fan},i}}{\dot{V}_{\text{rated},i}})^{1.5}\times W_{\text{rated},i},
\end{equation}
where $\dot{V}_{\text{fan},i}$ and $\dot{V}_{\text{rated},i}$ are the actual and rated airflow rates of \textit{i}'th FCU while $W_{\text{rated},i}$ is the fan's electrical power at rated conditions. Note that for typical constant-air-volume or three-speed FCUs, the exponent is commonly reduced to 1.5 in engineering practice; the present simulator retains this widely accepted approximation.

The water-side pump is modeled using a combination of manufacturer characteristic curves and affinity laws. At rated frequency, the pump head–flow relationship is represented by a quadratic polynomial \cite{ashrae2021}:
\begin{equation}
\label{eq:pump_curve}
\Delta P_{\rm{rated}} = \alpha_1 \dot{V}_{\rm{pump}}^2 + \alpha_2 \dot{V}_{\rm{pump}} + \alpha_3,
\end{equation}
where $\dot{V}_{\rm{pump}}$ is the volumetric flow rate and $\alpha_1$, $\alpha_2$, $\alpha_3$ are regression coefficients.

When the pump operates at variable speed, the head scales with the square of the speed ratio \cite{ashrae2021}:
\begin{equation}
\Delta P = \left( \frac{f_{\text{pump}}}{f_{\rm{rated}}} \right)^2 \Delta P_{\rm{rated}},
\end{equation}
where $f_{\text{pump}}$ and $f_{\rm{rated}}$ are the actual and rated motor frequencies. This relationship is embedded within the pipe network solver to determine the operating point that satisfies both the pump curve and the system resistance curve. Convergence of the iterative hydraulic solution achieved when:
\begin{equation}
\left| \left( \frac{f_{\text{pump}}}{f_{\rm{rated}}} \right)^2 \Delta P_{\rm{rated}} - (\alpha_1 \dot{V}_{\rm{pump}}^2 + \alpha_2 \dot{V}_{\rm{pump}} + \alpha_3) \right| \leq \varepsilon,
\end{equation}
with $\varepsilon$ being a small tolerance (typically $10^{-3} \sim 10^{-2}$ kPa).

Pump electrical power follows the cube-law scaling:
\begin{equation}
W_{\rm{pump}} = \left( \frac{f_{\text{pump}}}{f_{\rm{rated}}} \right)^3 W_{\rm{pump,rated}}.
\end{equation}

The water pipe network serves as the hydraulic and thermal backbone that links all water-side HVAC components, enabling fully coupled calculations of flow distribution, pressure losses, and temperature propagation through an iterative solver.

The network topology is formulated as a directed graph $\mathcal{G} = (\mathcal{V}, \mathcal{E})$, where $\mathcal{V}$ denotes the set of pipe branching/intersection nodes and $\mathcal{E}$ the set of directed edges between two nodes. The connectivity is compactly represented by the incidence matrix $\mathbf{M} \in \mathbb{R}^{|\mathcal{V}| \times |\mathcal{E}|}$, whose elements are defined as:
\begin{equation}
m_{i,j} =
\begin{cases}
\phantom{-}1,  &\text{if node $i$ is the upstream endpoint of branch $j$},\\
-1,  &\text{if node $i$ is the downstream endpoint of branch $j$},\\
\phantom{-}0, &\text{otherwise}.
\end{cases}
\end{equation}

At each simulation time step, the water volumetric flow rate $\dot{V}_{w,j}$ in every branch $j \in \mathcal{E}$ is obtained by solving the complete set of nonlinear hydraulic equations using the Newton--Raphson method. The variable-speed pump characteristic is directly incorporated as an implicit boundary condition, guaranteeing that mass continuity is satisfied at all nodes while simultaneously matching the system pressure--flow operating point.

After hydraulic convergence, the energy transport calculation is performed sequentially in the direction of flow. The outlet water temperature from each FCU coil is determined from the instantaneous zone cooling load and the coil entering water temperature using standard effectiveness--NTU relations. Temperature changes along pipe segments are computed by considering both convective transport and minor heat losses/gains to the surroundings (usually negligible for well-insulated lines). The updated nodal temperatures and branch flow rates are then passed back to the FCU models and pump model to close the coupled solution loop for the current time step.

This simulator serves as a testbed for evaluating the proposed control strategies, enabling the optimization of HVAC operation to minimize energy consumption while maintaining occupant comfort.

\section{Experimental Setting}
\label{sec:exp setting}

\subsection{Case Study Modeling}
\label{subsec:case_study_setup}

\begin{figure}
    \centerline{
    \includegraphics[width=\columnwidth, trim=0cm 0cm 0cm 0cm, clip]{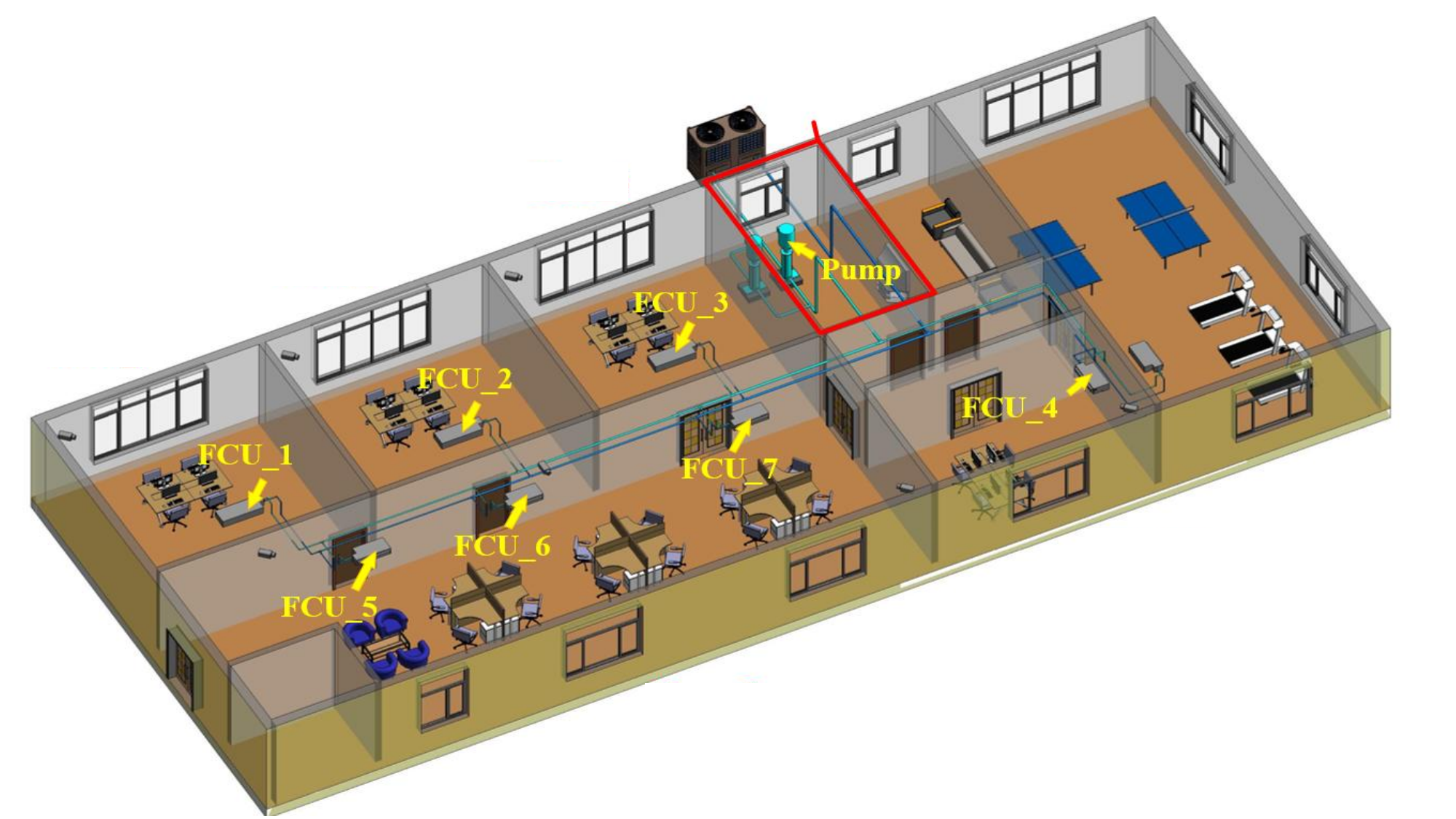}
    }    
    \caption{Layout of the office building \cite{ZHONG2025116219}.}   
    \label{fig:layout}
\end{figure}

The case study considers an office building in Hebei Province, China~\cite{xing2022honeycomb}. The building comprises a seven-zone office space, as shown in Fig.~\ref{fig:layout}, with one fan coil unit (FCU) installed in each thermal zone. This subsection describes the building layout, HVAC configuration, sensing infrastructure, and case-specific operating context used in the simulator-based evaluation.

\begin{figure*}
    \centering
    \includegraphics[width=2\columnwidth, trim=0cm 0cm 0cm 0cm, clip]{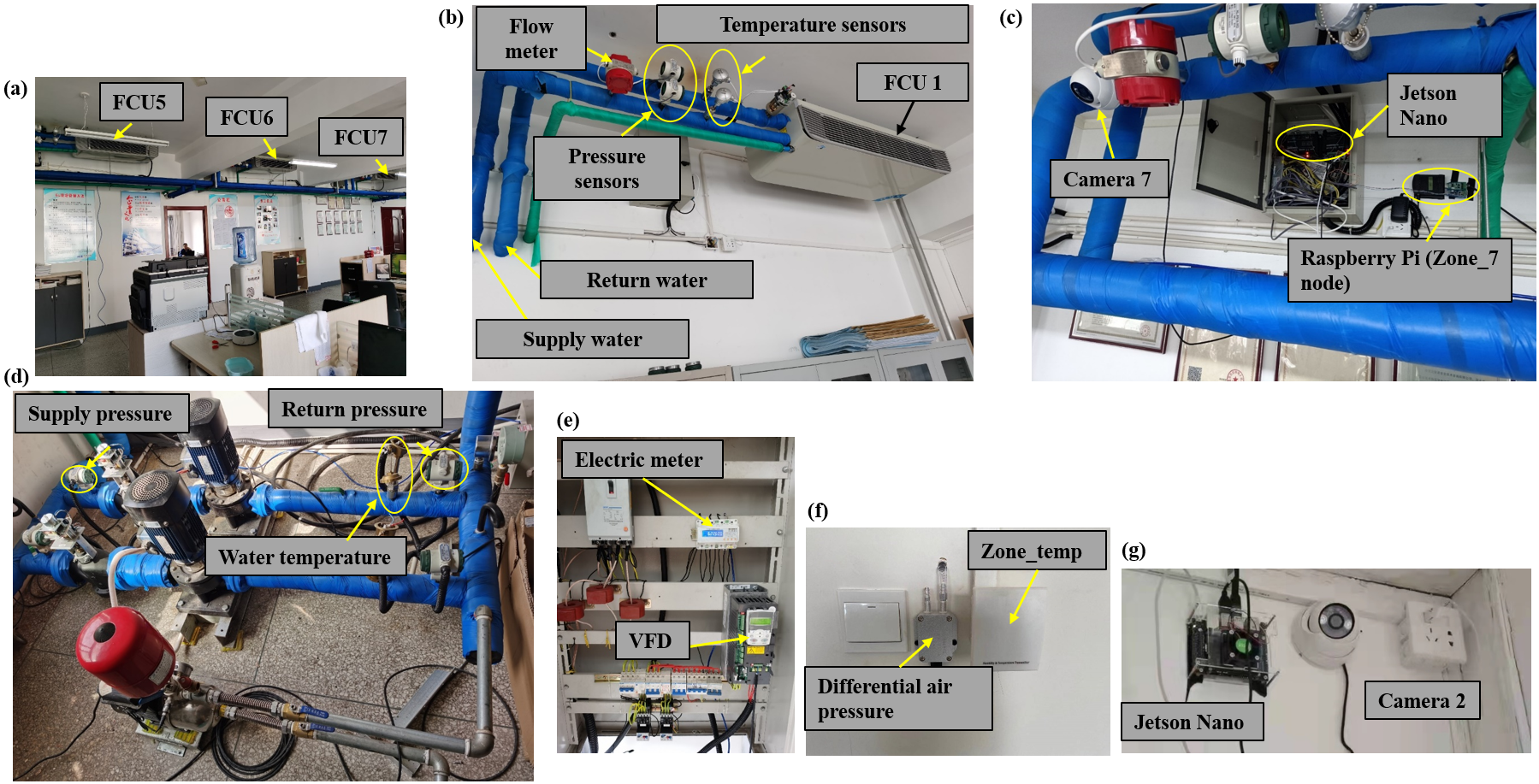}
    \caption{HVAC system configuration. (a) FCUs of zones 5 to 7. (b) Sensors of FCU 1. (c) Video sampling of zone 7. (d) Sensors of water pumps. (e) Electrical cabinet. (f) Temperature sensor. (g) Video camera of zone 2 \cite{ZHONG2025116219}.}
    \label{fig:real-hvac}
\end{figure*}

The building consists of several individual rooms and a hall. FCUs~1--4 each serve an individual cellular office, whereas FCUs~5--7 serve zones that allow inter-zone heat exchange. The FCUs receive chilled or heated water from a central refrigeration station comprising a heat pump and a circulating water pump. The pump distributes water to all FCUs through a closed-loop pipe network, while the heat pump provides the thermal source for cooling and heating. An overview of the HVAC system configuration and representative devices is shown in Fig.~\ref{fig:real-hvac}. The building adopts an FCU-based hydronic HVAC system serving multiple thermal zones, with zones 5--7 illustrated as representative examples in Fig.~\ref{fig:real-hvac}(a).

To monitor indoor thermal conditions and system operation, a sensing infrastructure is deployed throughout the building. At the zone level, each thermal zone is equipped with at least one indoor air temperature sensor for real-time monitoring and data collection, as exemplified in Fig.~\ref{fig:real-hvac}(f). 
Occupancy information is inferred from ceiling-mounted video cameras; Fig.~\ref{fig:real-hvac}(c) shows zone 7, and Fig.~\ref{fig:real-hvac}(g) shows zone 2.
At the terminal side, each FCU is instrumented with a water flow meter, supply and return water temperature sensors, and pressure sensors to capture coil-level thermal and hydraulic states (Fig.~\ref{fig:real-hvac}(b)). In the central plant, the circulating water pumps are equipped with inlet and outlet pressure sensors and return water temperature sensors to characterize system-level hydronic conditions (Fig.~\ref{fig:real-hvac}(d)). The pumps are driven by variable frequency drives (VFDs) with a control resolution of 1~Hz, and their electrical power consumption is measured via electrical meters installed in the electrical cabinet (Fig.~\ref{fig:real-hvac}(e)). Additional measurements include outdoor air temperature and system-level operational data. All sensors operated continuously during HVAC runtime throughout the study period. These measurements are used to characterize the building's thermal behavior and to support the simulator calibration described in Section~\ref{sec:simulator}.

The building operates primarily during standard working hours (9:00--19:00 on weekdays), with reduced occupancy around lunchtime. An average of 12.4 people occupied the office during the study period, with typical weekday usage and no regular weekend activity. The historical data were collected between August 7 and August 22, 2021~\cite{xing2022honeycomb}.

\subsection{Performance Evaluation Metrics}
\label{sec:metrics}

To comprehensively assess the proposed building control strategies, we adopt a dual-objective evaluation framework that quantifies both occupant thermal comfort and HVAC energy use. These metrics define the reinforcement learning (RL) reward signal and are also used for controller performance evaluation. This subsection details the thermal discomfort indices, energy metric, and composite reward function.

\subsubsection{Thermal Discomfort Metrics}

Thermal comfort is evaluated using the Predicted Mean Vote (PMV) and Predicted Percentage of Dissatisfied (PPD) indices, as introduced by Fanger~\cite{fanger1970thermal}. These indices provide a physiologically grounded assessment of human thermal sensation in indoor environments.

The PMV is calculated based on six parameters: air temperature ($T_a$), mean radiant temperature ($T_r$), relative humidity (RH), air velocity ($v$), clothing insulation ($I_{cl}$), and metabolic rate ($M$). Under the typical office conditions of the case study building, the following parameters are fixed in the PMV calculation: $v = 0.15$~m/s, RH $= 40\%$, $I_{cl} = 0.63$~clo, and $M = 1.1$~met. The mean radiant temperature $T_r$ is approximated as equal to the air temperature $T_a$ due to the well-insulated envelope and the limited radiant asymmetry in the office zones. Under these assumptions, PMV and the corresponding PPD are computed using the standard Fanger formulation.

A PMV value of 0 indicates neutral thermal sensation, with an acceptable range of $[-0.5, +0.5]$ as specified by ASHRAE Standard 55~\cite{ashrae2023standard55}, EN 15251~\cite{en200715251}, and ISO 7730~\cite{international1994iso}. The PPD, derived from PMV, predicts the percentage of occupants dissatisfied with the thermal environment. An acceptable PPD is typically below 10\%, corresponding to the PMV bounds above. In addition to PPD-based reward evaluation, the episode-level mean absolute PMV is reported as a supplementary comfort indicator in Section~\ref{sec:results}.

For multi-zone evaluation, we compute the zone-specific $\mathrm{PPD}_{i,t}$ at time step $t$ and aggregate it as the occupancy-weighted mean PPD across zones:
\begin{equation}
    \mathrm{PPD}_{\mathrm{mean},t} =
    \begin{cases}
    \dfrac{1}{N_t} \sum_{i=1}^{J} n_{i,t} \cdot \mathrm{PPD}_{i,t}, & N_t > 0, \\
    0, & N_t = 0,
    \end{cases}
\end{equation}
where $J=7$ is the number of zones, $n_{i,t}$ is the occupancy count in zone $i$ at time step $t$, and $N_t = \sum_{i=1}^{J} n_{i,t}$ is the total number of occupants.

\subsubsection{Energy Consumption Metric}

Energy consumption is measured as the total HVAC electrical power $P_t$ at each time step $t$. In the simulator-based evaluation, $P_t$ denotes the total modeled electrical power of the HVAC system returned by the simulator, while the underlying component models are described in Section~\ref{sec:simulator}. Power is reported in kilowatts (kW), and cumulative energy over an episode is computed as
\begin{equation}
    E = \sum_t P_t \cdot \Delta t,
\end{equation}
where $\Delta t = 5$ minutes $= 5/60$ hours is the control timestep.

\subsubsection{Composite Reward Function}

The overall objective is formalized through a reward function that balances thermal discomfort and energy consumption, with larger reward values indicating better control performance:
\begin{equation}
    r_t =
    \begin{cases}
    - \mathrm{PPD}_{\mathrm{mean},t} - \lambda_P P_t, & \text{if } N_t > 0, \\
    - \lambda_P P_t, & \text{if } N_t = 0.
    \end{cases}
\end{equation}
The hyperparameter $\lambda_P > 0$ controls the trade-off between comfort and energy use: larger values place more weight on energy savings, while smaller values emphasize thermal comfort. This formulation removes the comfort penalty during unoccupied periods while retaining the energy term, consistent with the occupancy-aware control objective defined in Section~\ref{sec:problem_formulation}.

\section{Results and Discussion}
\label{sec:results}

This section evaluates the proposed hierarchical LLM--RL framework (Section~\ref{sec:method}) in the calibrated 7-zone HVAC simulator (Section~\ref{sec:simulator}) under the metrics defined in Section~\ref{sec:metrics}. 
The evaluation aims to answer four key research questions:
\begin{itemize}
    \item RQ1: How does the proposed framework perform compared with baseline control strategies for comfort improvement and energy saving?
    \item RQ2: How does the LLM-generated action mask affect downstream RL exploration efficiency and training stability?
    \item RQ3: Is direct LLM control alone sufficient for multi-zone HVAC control?
    \item RQ4: How efficiently can the LLM component be trained and deployed in the proposed framework?
\end{itemize}

To address these questions, we first compare the overall control performance of the proposed method with state-of-the-art baseline algorithms (Section ~\ref{subsec:q1_overall}). We then analyze the role of the LLM-generated action mask in reducing the action space and accelerating RL training(Section  ~\ref{subsec:q2_mask_for_rl}). Finally, we evaluate the efficiency and effectiveness of the LLM component (Section ~\ref{subsec:q3_direct_vs_mask}) and present a representative case study to illustrate how the proposed controller makes decisions in practical scenarios (Section ~\ref{subsec:case_study}).

\begin{table*}[h]
\renewcommand\arraystretch{1.3}
\caption{
Performance comparison of different control paradigms in terms of mean PPD, mean absolute PMV, and HVAC energy use.
}
\centering
\label{tab:overall_performance}
\begin{tabular}{l|lccc}
\toprule
\textbf{Category} & \textbf{Alg.} & \textbf{PPD Mean (\%) $\downarrow$} & \textbf{PMV Abs Mean $\downarrow$} & \textbf{Energy Use (kWh) $\downarrow$} \\
\hline
\multirow{3}{*}{Vanilla LLM} 
& Meta-Llama-3-8B  & $27.72 \pm 2.07$ & $0.97 \pm 0.06$ & $134.46 \pm 2.28$ \\
& Qwen2.5-14B      & $16.41 \pm 0.84$ & $0.62 \pm 0.02$ & $156.23 \pm 1.26$ \\
& Qwen2.5-72B      & $15.56 \pm 0.89$ & $0.58 \pm 0.02$ & $157.42 \pm 2.02$ \\
\hline
\multirow{3}{*}{Vanilla RL}
& A2C   & $14.25 \pm 0.86$ & $0.48 \pm 0.03$ & $161.38 \pm 9.58$ \\
& PPO   & $13.59 \pm 0.44$ & $0.45 \pm 0.02$ & $152.53 \pm 3.08$ \\
& DQN   & $11.98 \pm 0.43$ & $0.39 \pm 0.03$ & $157.13 \pm 3.65$ \\
\hline
Hybrid
& LLM-RL (ours) & $\mathbf{7.30 \pm 0.62}$ & $\mathbf{0.25 \pm 0.02}$ & $140.90 \pm 2.28$ \\
\bottomrule
\end{tabular}
\end{table*}

\subsection{Overall Performance Evaluation}
\label{subsec:q1_overall}

Table~\ref{tab:overall_performance} compares the overall performance of vanilla LLM control, vanilla RL control, and the proposed hierarchical LLM--RL framework. Compared to baselines, the proposed framework delivers the strongest comfort performance while maintaining a favorable overall comfort--energy trade-off. 


\subsubsection{Occupant Comfort}
\label{subsubsec:q1_comfort}

The proposed LLM--RL framework achieves a mean PPD of $\mathbf{7.30\%}$ and a mean absolute PMV of $\mathbf{0.25}$ (Table~\ref{tab:overall_performance}), substantially outperforming both vanilla RL and vanilla LLM baselines.
Notably, the achieved mean PPD is below the widely used $10\%$ acceptability criterion implied by the PMV-based comfort standards (Section~\ref{sec:metrics}), indicating that the learned policy maintains a comfortable indoor environment on average.

Compared to the best vanilla RL baseline in terms of comfort, namely DQN, our framework reduces mean PPD from $11.98\%$ to $7.30\%$, corresponding to a relative reduction of $39.1\%$.
For reference, compared to PPO, the mean PPD is reduced from $13.59\%$ to $7.30\%$, corresponding to a $46.3\%$ relative reduction.
Compared with the strongest vanilla LLM baseline (Qwen2.5-72B), the mean PPD is reduced from $15.56\%$ to $7.30\%$ (a $53.1\%$ relative reduction), highlighting that direct LLM-based control is not sufficient for fine-grained comfort regulation under coupled thermal dynamics.

\subsubsection{Energy Consumption}
\label{subsubsec:q1_energy}

In addition to improving comfort, the proposed framework also achieves a strong energy-saving performance.
In Table~\ref{tab:overall_performance}, our method consumes $140.90$~kWh, which is lower than all vanilla RL baselines, including PPO ($152.53$~kWh), DQN ($157.13$~kWh), and A2C ($161.38$~kWh).
It also consumes less energy than the stronger vanilla LLM baselines, namely Qwen2.5-14B ($156.23$~kWh) and Qwen2.5-72B ($157.42$~kWh).

Although Direct LLM (Meta-Llama-3-8B) achieves a lower energy value of $134.46$~kWh, this comes at the cost of severe comfort degradation ($27.72\%$ mean PPD), indicating under-conditioning rather than a better comfort-energy trade-off.
We therefore interpret the proposed framework as achieving a more desirable balance between occupant comfort and energy efficiency, rather than simply minimizing energy consumption.

We attribute this energy advantage to the division of labor between the two layers: the LLM provides a feasible action subset that avoids obviously wasteful behaviors (e.g., strong cooling in persistently unoccupied zones), and the RL layer performs fine-grained selection within that subset to avoid unnecessary intensity or oscillations.

\subsection{Impact of LLM-guided Action Mask on RL Training}
\label{subsec:q2_mask_for_rl}

This section investigates the role of the LLM-generated action mask in assisting RL training.
While RL can in principle optimize long-horizon objectives, vanilla RL becomes inefficient and unstable in multi-zone HVAC systems because of the exponential growth of the joint action space and the prevalence of low-quality exploratory actions.
In our 7-zone setting, each FCU has four discrete fan-speed levels, $\{0,1,2,3\}$, yielding a full joint action space of $|\mathcal{A}|=4^7=16384$.
This combinatorial branching factor substantially increases the exploration burden of RL, especially in the early training stage when behavior remains highly stochastic.

Our framework addresses this limitation by using the fine-tuned LLM (Section~\ref{sec:sft}) to generate state-dependent feasible action masks, thereby restricting RL exploration to a semantically meaningful subset $\widetilde{\mathcal{A}}(s_t)\subset\mathcal{A}$ (Section~\ref{sec:masked_dqn}).
We first establish the overall end-to-end training advantage of LLM guidance and then explain this advantage from two mechanism-oriented perspectives: action-space reduction and the quality of the retained masked actions.

\subsubsection{Training Dynamics and Final Performance Across RL Algorithms}
\label{subsubsec:q2_convergence}

\begin{figure}[t]
    \centering
    \includegraphics[width=\columnwidth]{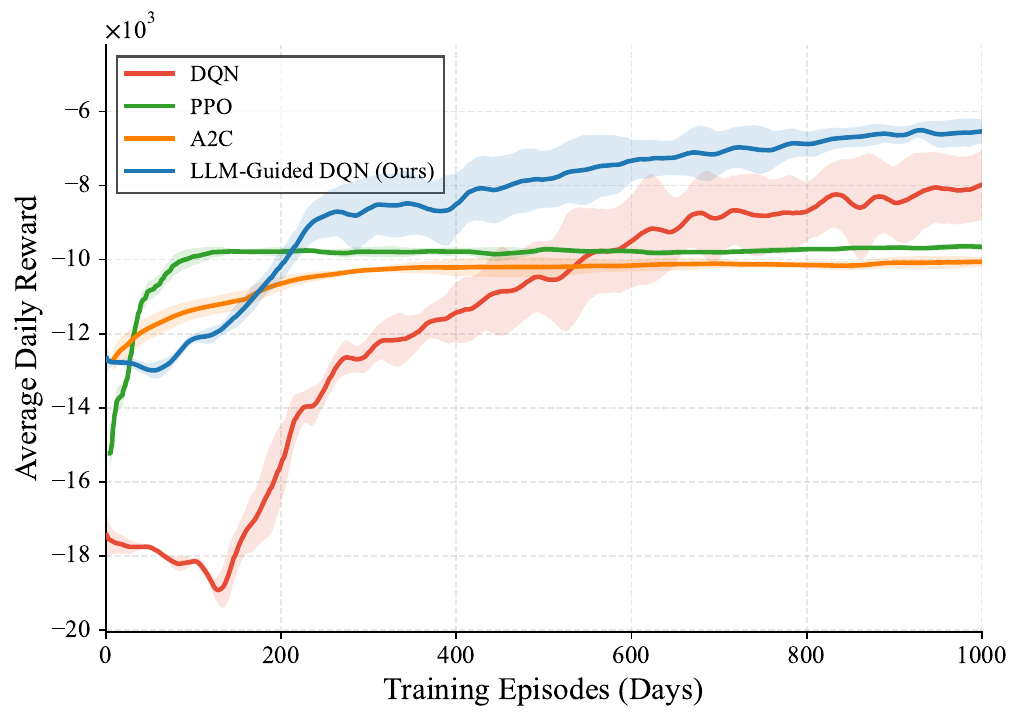}
    \caption{Training curves of A2C, PPO, DQN, and LLM-guided DQN (ours). The x-axis shows training episodes, where each episode corresponds to one natural day. Curves denote smoothed mean daily reward, and shaded regions indicate one standard deviation across five random seeds. LLM-guided DQN achieves the best final reward, the best peak reward, and the largest (least negative) AUC, indicating both faster convergence and better final performance. All results are averaged over five random seeds.}
    \label{fig:rl_training_curve}
\end{figure}

We compare the full training behaviors of all RL algorithms, including A2C, PPO, DQN, and our LLM-guided DQN.
Fig.~\ref{fig:rl_training_curve} reports the training curves in terms of average daily reward versus training episodes, where each episode corresponds to one natural day.
Because rewards in this task are negative, values closer to zero indicate better control performance.

Among all methods, LLM-guided DQN achieves the best final training performance.
In the final stage of training, it reaches a reward of $-6533.3$, outperforming DQN ($-7886.5$), PPO ($-9655.2$), and A2C ($-10088.0$).
In particular, relative to the strongest baseline in terms of final reward (DQN), our method yields a 17.16\% improvement.
The proposed method also achieves the best peak reward during training ($-6506.3$), whereas DQN, PPO, and A2C peak at $-7973.6$, $-9640.3$, and $-10061.3$, respectively.

Beyond final performance, Fig.~\ref{fig:rl_training_curve} also shows a clear advantage in convergence speed.
Using the area under the learning curve (AUC) as a summary of overall training efficiency, LLM-guided DQN attains the largest (i.e., least negative) AUC, $-8.48\times10^6$, compared with $-1.16\times10^7$ for DQN, $-9.88\times10^6$ for PPO, and $-1.03\times10^7$ for A2C.
This indicates that the proposed method reaches higher-reward regimes earlier and maintains better performance throughout training, rather than merely obtaining a better endpoint.
Relative to vanilla DQN, the terminal cross-seed standard deviation is also reduced from 812.2 to 256.0, suggesting that the LLM-guided action mask improves not only sample efficiency but also the stability of value-based training.
Although PPO and A2C exhibit smaller terminal dispersion, they converge to substantially worse reward plateaus.

Overall, these results establish that LLM-guided masking improves both training efficiency and final policy quality at the system level.
These training-curve observations are also consistent with Table~\ref{tab:overall_performance}, indicating that faster convergence is achieved together with stronger overall control performance.
To understand where this advantage comes from, the next two subsections examine how strongly the mask compresses the original joint action space and whether the retained actions are intrinsically higher quality.

\subsubsection{Action Space Reduction}
\label{subsubsec:q2_reduction}

To explain the observed training advantage, we first quantify how strongly the LLM-generated mask compresses the original joint action space.
At each control step $t$, the fine-tuned LLM predicts a feasible fan-speed set $M_i(s_t)\subseteq\{0,1,2,3\}$ for each zone $i$.
Accordingly, the size of the valid joint action space is
\begin{equation}
|\widetilde{\mathcal{A}}(s_t)|=\prod_{i=1}^{7}|M_i(s_t)|.
\end{equation}

To quantify the reduction effect in a normalized form, we define the remaining action-space percentage at time step $t$ as
\begin{equation}
P_t = \frac{|\widetilde{\mathcal{A}}(s_t)|}{|\mathcal{A}|}\times 100\%,
\label{eq:remaining_action_percentage}
\end{equation}
where a smaller $P_t$ indicates a stronger pruning effect and a more compact search space for RL exploration.
Over one episode, the average remaining percentage is computed as
\begin{equation}
\bar{P}_{\mathrm{epi}}=\frac{1}{T}\sum_{t=1}^{T}P_t,
\label{eq:mean_remaining_action_percentage}
\end{equation}
where $T$ denotes the total number of control steps in the episode.

\begin{figure*}[t]
    \centering
    \includegraphics[width=\textwidth]{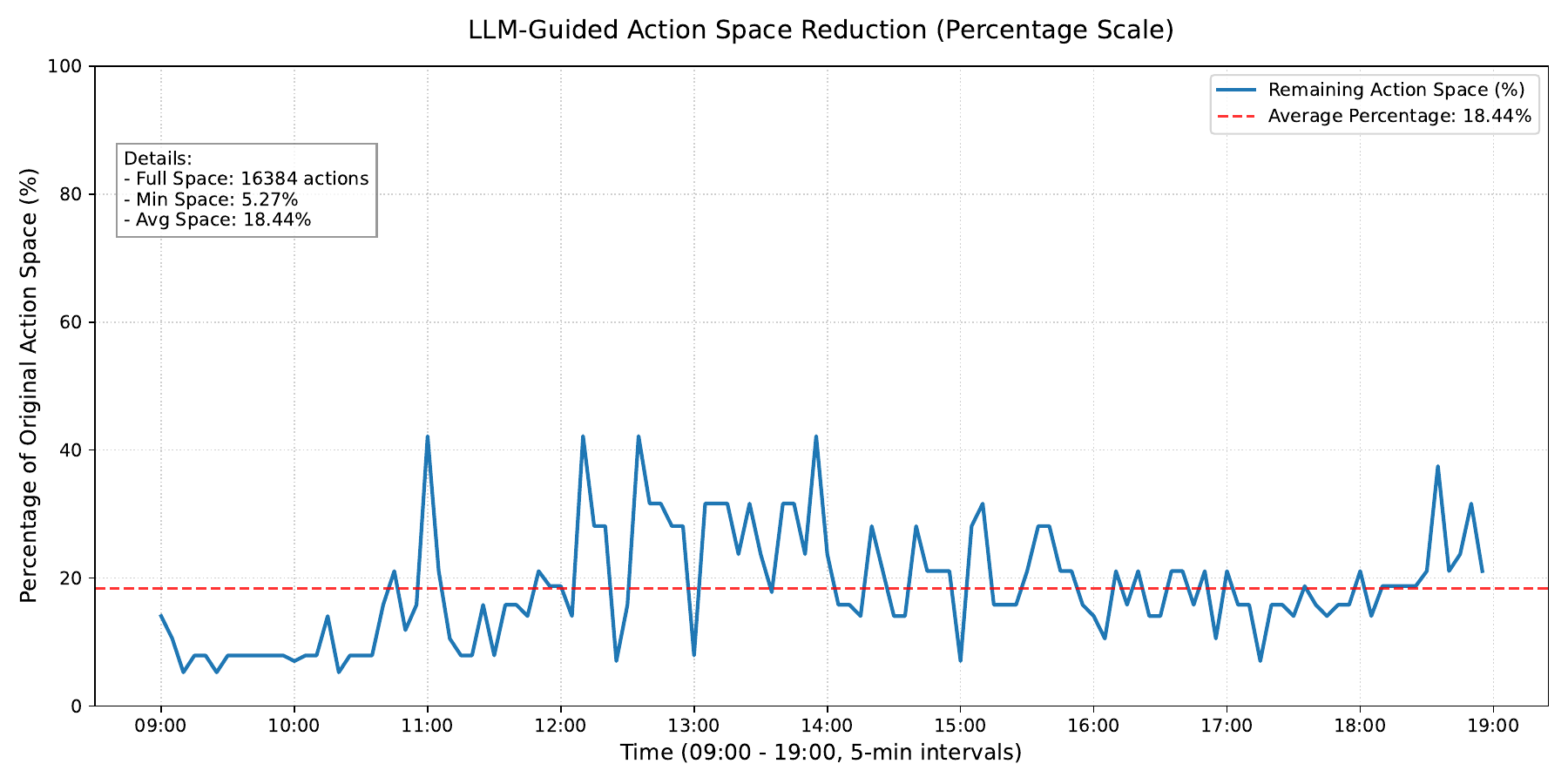}
    \caption{Temporal variation of the remaining action-space percentage over a representative episode.
    The y-axis shows the percentage of the original joint action space retained after LLM-guided masking, as defined in Eq.~\eqref{eq:remaining_action_percentage}.
    The x-axis represents the real time within one working day (09:00--19:00, with a 5-minute control interval).
    The dashed horizontal line denotes the episode-wise average remaining percentage.}
    \label{fig:action_space_reduction}
\end{figure*}

Fig.~\ref{fig:action_space_reduction} illustrates the temporal evolution of $P_t$ over a representative episode. It can be seen that the remaining action-space percentage stays consistently far below the full action space throughout the day. It indicates that the LLM-based mask continuously compresses the original action space into a much smaller feasible subset.
Meanwhile, the curve varies with operating conditions rather than remaining constant.
During relatively stable periods, the percentage becomes lower, suggesting that more redundant actions can be safely pruned.
In contrast, during transition periods associated with dynamic changes in occupancy or thermal load, the percentage increases moderately, indicating that a larger feasible set is preserved to maintain control flexibility.
This result shows that the mask is not only restrictive but also adaptive to time-varying system conditions.

Table~\ref{tab:action_space_percentage} further summarizes the key statistics of the remaining action-space percentage over the same episode.
The maximum remaining proportion is $42.19\%$, corresponding to 6912 valid joint actions, whereas the minimum drops to only $5.27\%$, or 864 actions.
On average, the mask retains $18.44\%$ of the original action space, equivalent to 3021.30 valid joint actions.
In other words, approximately $81.56\%$ of the full combinatorial action space is eliminated on average, substantially reducing the exploration burden for the downstream RL agent.

\begin{table}[t]
\centering
\caption{Summary statistics of the remaining action-space percentage over a representative episode.
Percentages are reported relative to the original joint action space of size $16384$, and the corresponding valid-action counts are provided for reference.}
\label{tab:action_space_percentage}
\begin{tabular}{lcc}
\hline
\textbf{Statistic} & \textbf{Percentage} & \textbf{Valid Actions} \\
\hline
Maximum & 42.19\% & 6912 \\
Minimum & 5.27\%  & 864 \\
Average & 18.44\% & 3021.30 \\
\hline
\end{tabular}
\end{table}

Overall, these results demonstrate that the LLM-generated mask functions not merely as a heuristic filter, but as an effective action-space reduction mechanism.
By projecting the original joint action space onto a compact and state-dependent feasible subset, the mask directly reduces the exploration burden faced by the RL agent and provides one mechanistic explanation for the stronger convergence behavior observed in Fig.~\ref{fig:rl_training_curve}.
However, action-space compression alone does not guarantee better learning; the retained actions must also be of sufficiently high quality.
This issue is examined next.

\subsubsection{Cold Start and Training Stage}
\label{subsubsec:q2_coldstart}

The benefit of action-space reduction depends not only on shrinking the feasible set, but also on whether the retained actions are high-quality.
A useful mask should exclude actions that are operationally implausible or consistently harmful to the comfort-energy trade-off.
To examine this point, we further analyse LLM-guided masking from two complementary perspectives: its immediate effect at cold start and its downstream effect during end-to-end RL training.

\noindent\textbf{Cold start.}
We first consider a cold-start regime in which action selection is purely random.
Specifically, we compare:
\begin{itemize}
    \item \textbf{Full-Random:} uniformly sample $a_t \sim \mathcal{A}$.
    \item \textbf{Masked-Random:} query the LLM to obtain $\widetilde{\mathcal{A}}(s_t)$ and uniformly sample $a_t \sim \widetilde{\mathcal{A}}(s_t)$.
\end{itemize}
Since neither policy is updated using reward feedback, any performance difference in this setting can be attributed to the intrinsic quality of the LLM-generated feasible set itself.

\begin{figure*}[t]
    \centering
    \includegraphics[width=2\columnwidth]{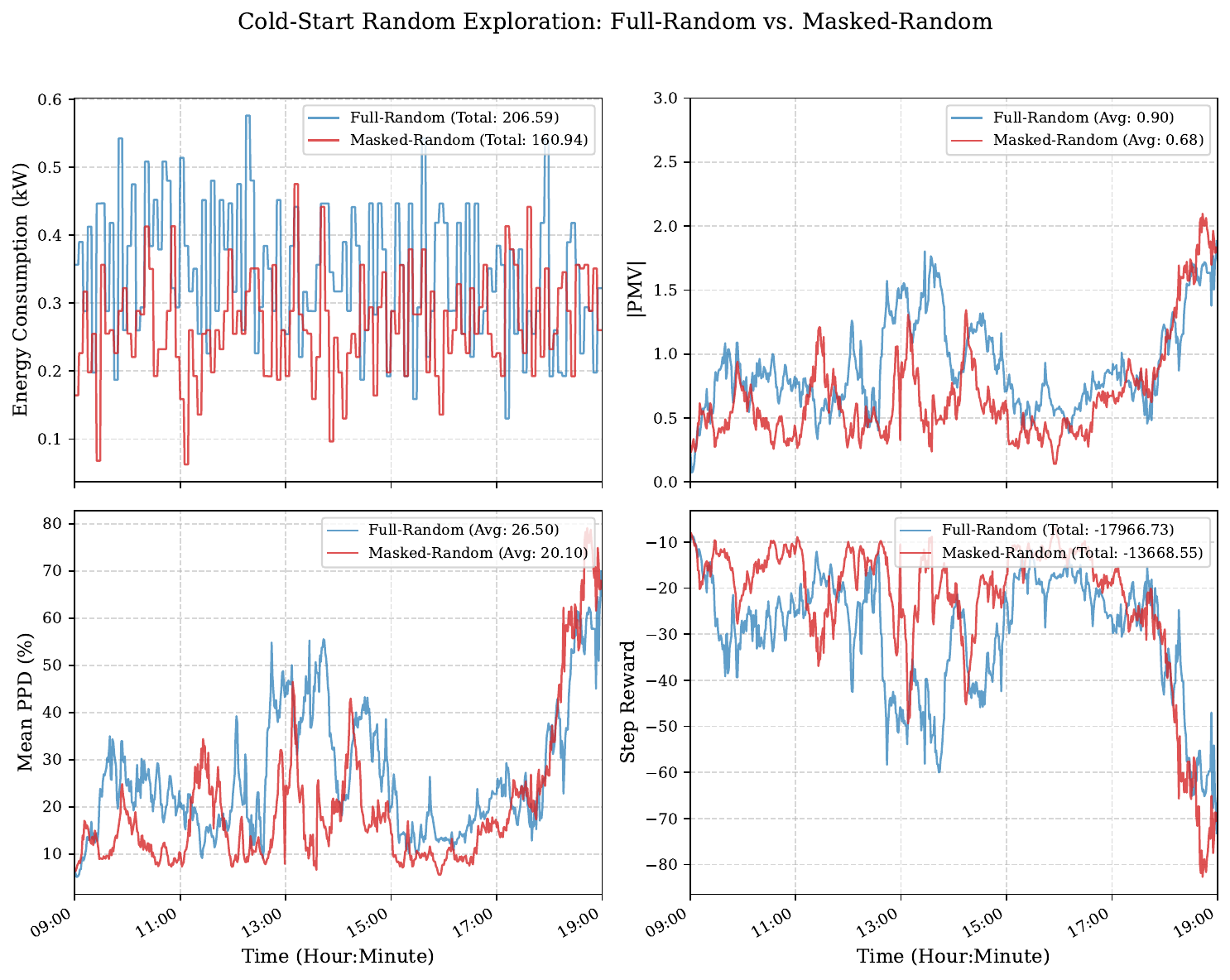}
    \caption{Time-resolved cold-start comparison under purely random exploration on a representative control day.
    Full-Random samples uniformly from the full action space $\mathcal{A}$, whereas Masked-Random samples uniformly from the LLM-feasible set $\widetilde{\mathcal{A}}(s_t)$.
    Compared with Full-Random, Masked-Random reduces total energy by 22.10\%, average $|\mathrm{PMV}|$ by 24.51\%, and average PPD by 24.16\%, while improving total reward by 23.92\%.}
    \label{fig:cold_start_random}
\end{figure*}

Fig.~\ref{fig:cold_start_random} presents a time-resolved comparison on a representative control day from 09:00 to 19:00, including energy consumption, absolute PMV, mean PPD, and step reward.
Even under purely random exploration, Masked-Random achieves a substantially better comfort--energy trade-off than Full-Random.
Compared with Full-Random, Masked-Random reduces total energy from 206.59 to 160.94 (22.10\%), lowers average $|\mathrm{PMV}|$ from 0.90 to 0.68 (24.51\%), decreases average PPD from 26.50\% to 20.10\% (24.16\%), and improves total reward from $-17966.73$ to $-13668.55$ (23.92\%).
These gains are obtained before any reward-driven policy learning takes place, indicating that the LLM-generated mask already filters out a large fraction of low-quality actions at the very beginning of interaction.

\begin{figure}[h]
    \centering
    \includegraphics[width=\columnwidth]{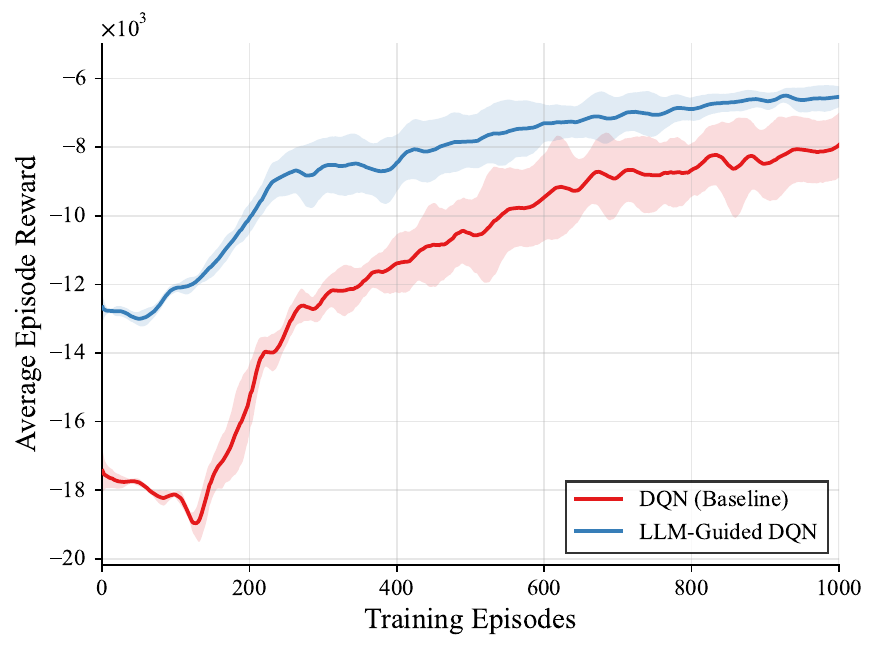}
    \caption{Learning curves of DQN and LLM-guided DQN under identical training budgets.
    Curves show smoothed mean evaluation return across random seeds, and shaded regions indicate one standard deviation.
    Averaged over the last 5\% of training, LLM-guided DQN improves the final evaluation return from $-8060.20$ to $-6565.14$ (18.55\%), while also exhibiting substantially smaller final cross-seed dispersion (256.04 vs.\ 812.22).}
    \label{fig:dqn_vs_masked_dqn}
\end{figure}

\noindent\textbf{Training dynamics.}
We next evaluate whether this cold-start advantage translates into stronger performance during full RL training.
Fig.~\ref{fig:dqn_vs_masked_dqn} compares DQN and LLM-guided DQN under identical training budgets and multiple random seeds. The LLM-guided DQN converges to a better final policy than the DQN baseline.
When averaging the raw evaluation returns over the last 5\% of training, LLM-guided DQN attains $-6565.14$, compared with $-8060.20$ for DQN, corresponding to an 18.55\% improvement.
Moreover, the cross-seed standard deviation at the final evaluation point decreases from 812.22 to 256.04, indicating substantially more stable convergence.

Taken together with the action-space reduction analysis and the broader multi-algorithm comparison in Fig.~\ref{fig:rl_training_curve}, these results suggest that LLM-guided masking is beneficial not merely because it reduces the number of admissible actions, but because it reshapes exploration toward a smaller and higher-quality subset of actions, thereby improving both the quality of collected experience and the stability of downstream learning.

\subsection{Efficiency and Effectiveness of the LLM Component}
\label{subsec:q3_direct_vs_mask}

This subsection examines the role of the LLM from three complementary perspectives: 
(i) whether the LLM should be used as a direct controller or as a generator of feasible action masks, 
(ii) how efficiently the mask generator can be deployed online, and 
(iii) whether the learned masks are reliable enough to support downstream RL. 
We begin with the most fundamental comparison: Vanilla LLM action generation versus the proposed \emph{LLM mask + RL refinement} paradigm.

\subsubsection{LLM scale v.s. hierarchical structure}

Under \emph{Vanilla LLM control}, the model outputs one discrete fan speed for each zone at every control step. 
By contrast, in the proposed hierarchical framework, the fine-tuned LLM outputs a feasible fan-speed set for each zone, and the masked DQN selects the final action by optimizing the learned $Q$-values within $\widetilde{\mathcal{A}}(s_t)$ according to Eq.~\eqref{eq:masked_argmax}. 
This comparison allows us to distinguish two possible sources of improvement: \emph{LLM scale} and \emph{hierarchical control structure}.

\begin{figure*}[t]
    \centering
    \includegraphics[width=\textwidth]{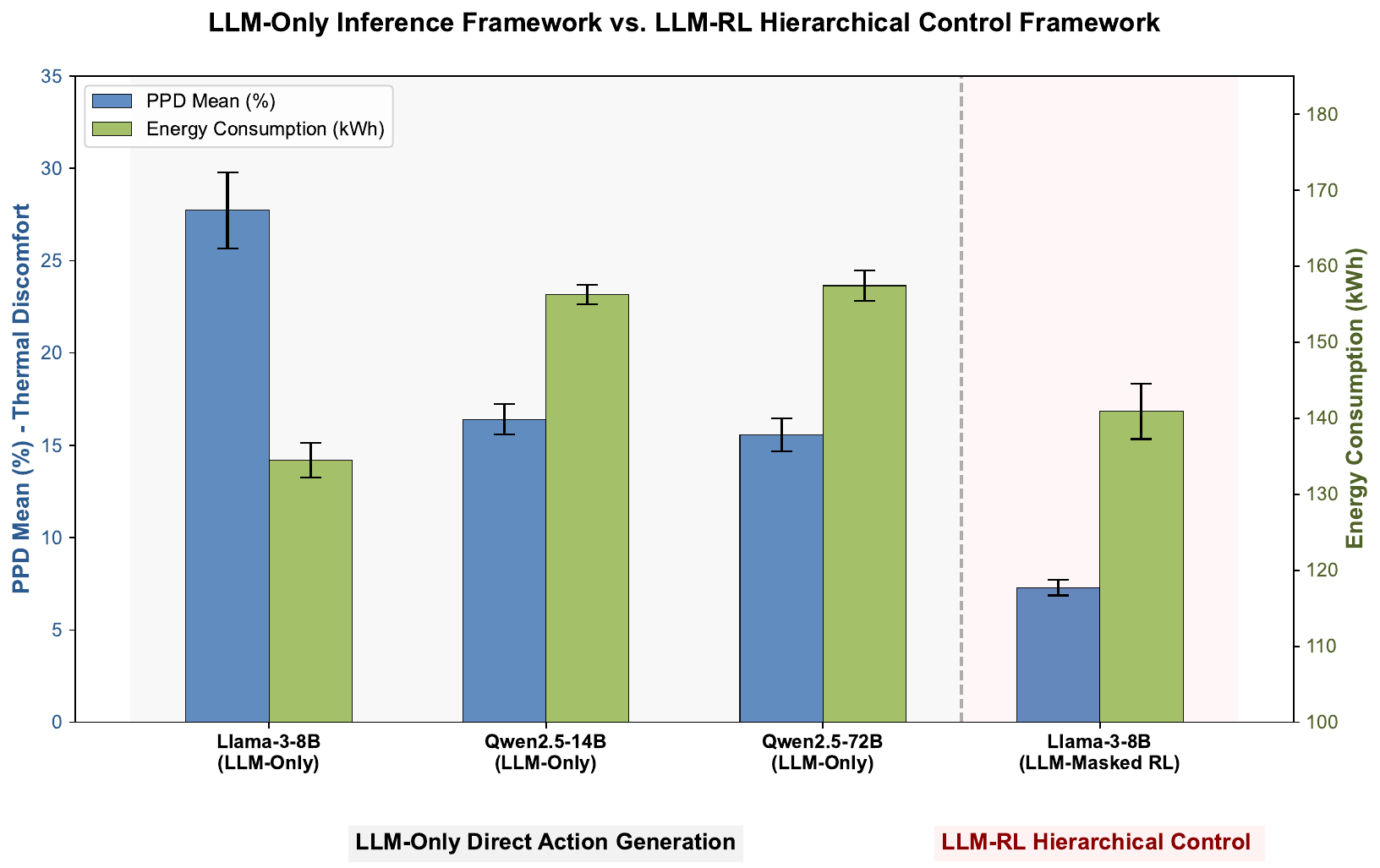}
    \caption{Comparison between direct LLM action generation and the hierarchical LLM-guided RL paradigm. Enlarging the LLM improves direct-control comfort, but also increases energy consumption. By contrast, the proposed hierarchical controller achieves the best comfort performance with a better overall comfort--energy trade-off.}
    \label{fig:q3_paradigm_compare}
\end{figure*}

Fig.~\ref{fig:q3_paradigm_compare} shows that enlarging the LLM under direct control does improve comfort: moving from Llama-3-8B to Qwen2.5-14B and Qwen2.5-72B consistently lowers PPD. 
However, this gain is accompanied by a clear increase in energy consumption, suggesting that larger LLMs tend to rely on more conservative cooling when they are required to generate final actions directly. 
In contrast, the hierarchical controller, although using only Llama-3-8B as the mask generator, achieves the best comfort performance while maintaining lower energy use than the stronger Vanilla LLM baselines. 
This indicates that the major benefit comes from the separation between \emph{semantic feasibility generation} and \emph{reward-driven action refinement}, rather than from scaling the language model alone.

\begin{figure}[t]
    \centering
    \includegraphics[width=\columnwidth]{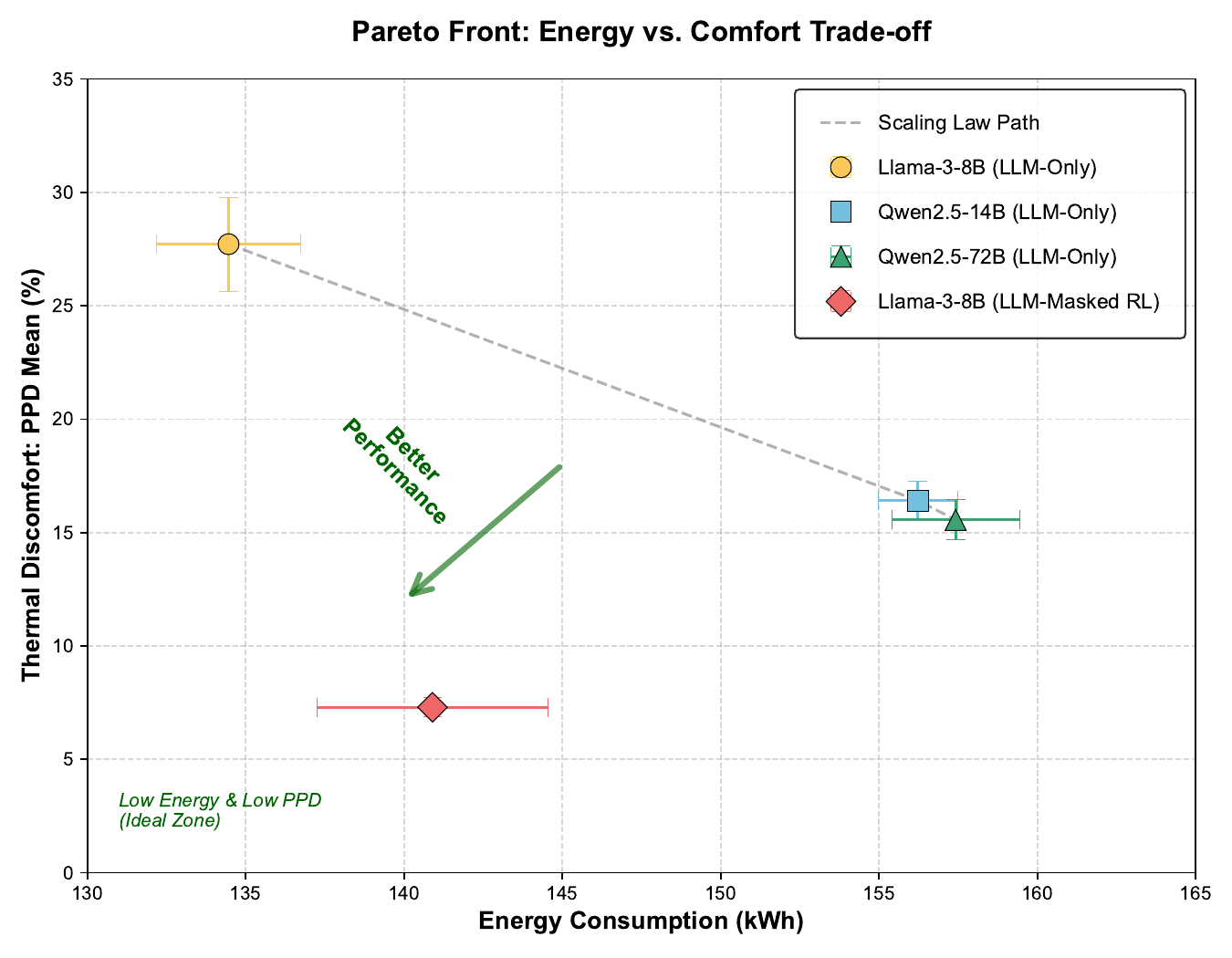}
    \caption{Pareto frontier of energy consumption and thermal discomfort. The direct-LLM baselines follow a scaling path toward better comfort but higher energy use, whereas the hierarchical controller moves closer to the lower-left ideal region.}
    \label{fig:q3_pareto_front}
\end{figure}

The same conclusion is reinforced by the Pareto view in Fig.~\ref{fig:q3_pareto_front}. 
The direct-LLM baselines form a clear scaling path in which improved comfort is obtained mainly by moving toward a higher-energy operating regime. 
The proposed hierarchical controller shifts the operating point toward the lower-left region of the energy--comfort plane, i.e., closer to the ideal zone of simultaneously low PPD and low energy use. 
Therefore, RL refinement does not simply polish the LLM output; it fundamentally changes the comfort-energy trade-off achieved by the controller.

Table~\ref{tab:overall_performance} reports the exact numerical results. 
Overall, these observations support a central design insight of this work: in multi-zone HVAC control, LLMs are more effective as generators of structured, human-aligned feasible sets, while RL should remain responsible for closed-loop optimization under delayed rewards. 
In this sense, \emph{hierarchical structure matters more than scale}.

\subsubsection{Inference efficiency: caching LLM masks for fast online control}
\label{subsubsec:q3_cache}

Although HVAC control operates at a relatively coarse and slow time scale (e.g., 10,30-minute control intervals), naively querying the LLM at every step can still introduce unnecessary inference time. 
To reduce the online computational cost, we adopt an \textbf{LLM + cache} strategy. 
Unlike a cold-start cache, the cache in our system is \emph{not initially empty}. 
Instead, it is constructed offline by running the fine-tuned LLM over a set of representative historical states collected from the building environment. 
These offline-generated state--mask pairs form an initial cache. 
During online control, if the current state (or its discretized/hash representation) matches a cached state, the system directly reuses the stored mask. 
For unseen states, the LLM is queried once and the resulting mask is then added to the cache for future reuse.

Because cache lookup uses a discretized state key rather than exact prompt matching, the cached controller may produce slightly different masks from step-wise LLM querying; thus, caching is an approximate acceleration mechanism rather than a strictly action-equivalent substitute. This design is well aligned with the strong thermal inertia and repeated occupancy patterns in buildings, where many control states recur across time steps and across days. 
Consequently, the majority of mask queries can be served directly from the cache.

To evaluate the practical benefit of caching, we compare two inference settings on a representative episode (seed 0) consisting of 120 control steps: 
(1) querying the LLM at every step (\textbf{No Cache}) and 
(2) using the proposed \textbf{LLM + Cache} mechanism. Table~\ref{tab:cache_efficiency} summarizes the efficiency statistics.
Without caching, each control step requires an LLM inference call, resulting in an average step latency of 2.27~s and a total episode runtime of 272.52~s. 
As a result, the average step latency drops to 0.087~s and the total runtime decreases to 10.46~s. 
This corresponds to a \textbf{96.16\% reduction} in both total runtime and average per-step latency.

The minimum latency of 0.011~s indicates that cache hits incur almost negligible overhead, whereas the occasional latency spikes (up to 2.39~s) correspond to cache misses that still require full LLM inference. 
Importantly, this efficiency improvement does not degrade control performance in the representative run: the episode reward with caching ($-5746.74$) is comparable to that obtained without caching ($-6158.01$). 
Given the 5-minute control interval adopted in this work, the resulting average inference latency of 0.087~s is negligible relative to the system dynamics, demonstrating that the proposed LLM-guided controller can be deployed in real-time building management systems without introducing practical computational bottlenecks.

\begin{table}[t]
\centering
\caption{Inference efficiency comparison between querying the LLM at every step and using the proposed LLM+Cache mechanism. Statistics are reported for a representative episode with 120 control steps.}
\label{tab:cache_efficiency}
\begin{tabular}{lcc}
\hline
\textbf{Metric} & \textbf{No Cache} & \textbf{LLM + Cache} \\
\hline
Total Episode Time (s) & 272.52 & 10.46 \\
Average Step Time (s) & 2.27 & 0.087 \\
Maximum Step Time (s) & 2.93 & 2.39 \\
Minimum Step Time (s) & 2.07 & 0.011 \\
Total Steps & 120 & 120 \\
Episode Reward & -6158.01 & -5746.74 \\
\hline
\end{tabular}
\end{table}

\subsubsection{Training efficiency}
\label{subsubsec:q3_sft_curve}

The LLM component is trained via supervised fine-tuning (SFT) with LoRA (Section~\ref{sec:sft}), which provides a parameter-efficient way to adapt the base model to the building-specific mask prediction task. 
In our implementation, Meta-Llama-3-8B-Instruct is fine-tuned on at most 1000 samples, with 10\% held out for validation. 
The input cutoff length is set to 2048 tokens. 
We use a per-device training batch size of 2 with 8 gradient accumulation steps (effective batch size 16), a learning rate of $2\times10^{-5}$, a cosine learning-rate schedule with 10 warmup steps, and 5 training epochs in FP16 precision. 
These settings constitute a lightweight SFT configuration while maintaining stable optimization.

\begin{table}[t]
\centering
\caption{Core hyperparameters for LoRA-based SFT of the LLM mask predictor.}
\label{tab:sft_core_config}
\begin{tabular}{lc}
\hline
\textbf{Parameter} & \textbf{Value} \\
\hline
Base model & Meta-Llama-3-8B-Instruct \\
Finetuning type & LoRA \\
Training samples & 1000 \\
Validation split & 10\% \\
Cutoff length & 2048 \\
Effective batch size & 16 \\
Learning rate & $2\times10^{-5}$ \\
LR scheduler & Cosine \\
Warmup steps & 10 \\
Training epochs & 5 \\
Precision & FP16 \\
\hline
\end{tabular}
\end{table}

\begin{figure*}[t]
    \centering
    \subfigure[Training loss during SFT.]{
        \includegraphics[width=0.48\textwidth]{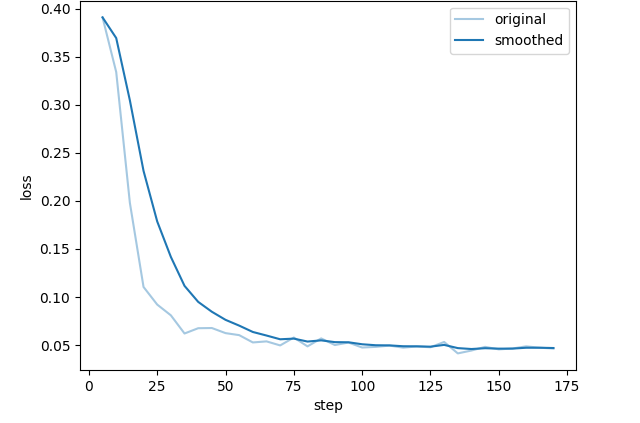}
    }
    \hfill
    \subfigure[Evaluation loss on the validation split.]{
        \includegraphics[width=0.48\textwidth]{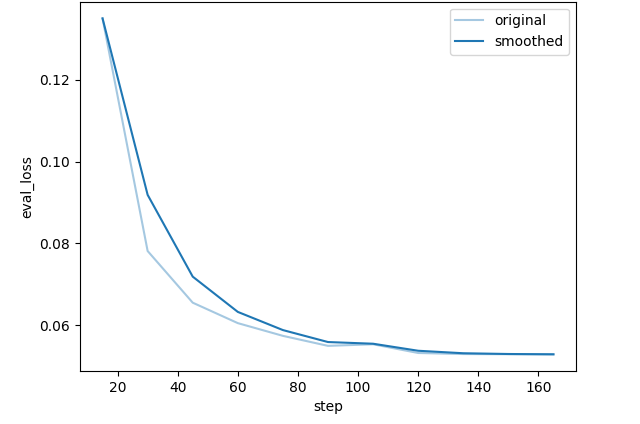}
    }
    \caption{Training dynamics of LoRA-based supervised fine-tuning for the LLM mask predictor. The left panel shows the training loss and the right panel shows the evaluation loss on the held-out validation split. Together, the two curves show that the model adapts efficiently to the building-specific mask prediction task under a lightweight fine-tuning setup.}
    \label{fig:sft_training_curve}
\end{figure*}

Fig.~\ref{fig:sft_training_curve} presents the training dynamics of the SFT stage. 
The training loss decreases and gradually stabilizes, while the evaluation loss on the held-out validation split remains well behaved without obvious divergence. 
This indicates that the LoRA-based adaptation converges efficiently under a limited-data setting and does not exhibit severe overfitting. 
Therefore, the LLM mask predictor can be trained with modest computational cost and then integrated into the hierarchical controller as a practical semantic mask generator.

\subsection{Representative case study: interpretable and smooth decision}
\label{subsec:case_study}

Aggregate metrics quantify the overall comfort--energy trade-off, but they do not fully reveal how the learned controller reacts to concrete operational events within a day. 
To provide a more interpretable view, we examine a representative workday trajectory of Room~6 under the proposed hierarchical LLM--RL controller. 
Room~6 is a particularly informative example because it belongs to the thermally coupled zone cluster (Rooms~5--7), where the controller must respond not only to local occupancy changes but also to inter-zone thermal interactions.

\begin{figure*}[t]
    \centering
    \includegraphics[width=\textwidth]{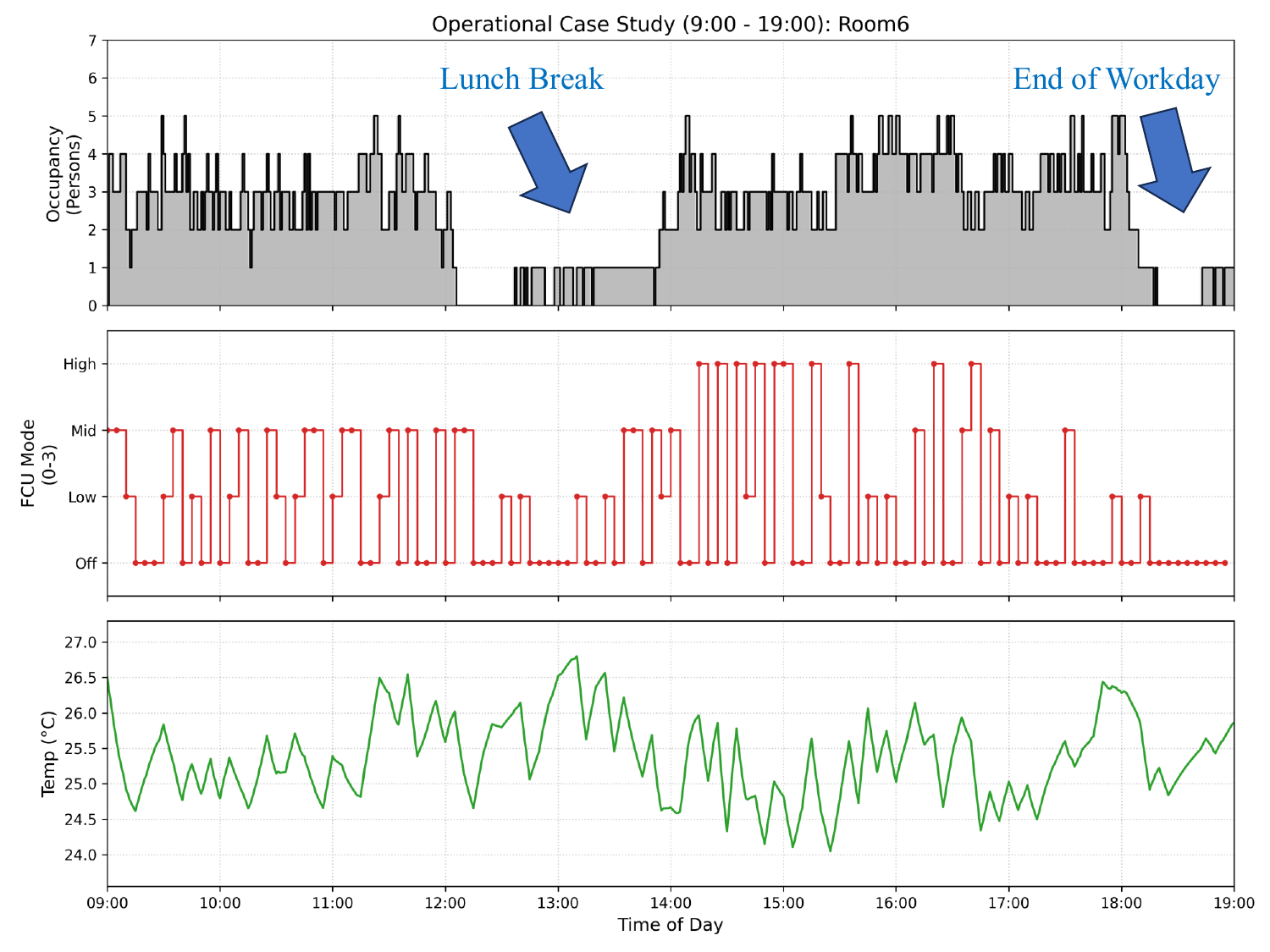}
    \caption{Representative-day control trajectory of Room~6 under the proposed hierarchical LLM--RL controller. From top to bottom: occupancy, FCU mode, and indoor temperature from 09:00 to 19:00. The lunch-break period (12:00--14:00) and the end-of-workday period (18:00--19:00) are highlighted. In both periods, the reduction in occupancy is accompanied by a corresponding reduction in FCU mode, illustrating occupancy-aware setback control.}
    \label{fig:case_room6}
\end{figure*}

Fig.~\ref{fig:case_room6} shows three synchronized signals over the working day: room occupancy, FCU mode, and indoor temperature. 
Several meaningful patterns can be observed. 

(1) During the regular occupied hours in the morning, Room~6 is typically occupied by about 2--4 persons, and the FCU mainly operates at Low or Mid mode, with stronger cooling invoked only when the thermal load becomes higher. 
The indoor temperature is maintained within a moderate range of roughly 24--27$^\circ$C, indicating that the controller does not rely on persistent high-speed actuation to maintain acceptable thermal conditions. 

(2) The lunch-break period (12:00--14:00) provides the clearest evidence of interpretable, occupant-centric control. 
As occupancy drops sharply to nearly zero, the FCU mode is correspondingly reduced to Off or Low for most of this interval. 
Instead of maintaining unnecessary cooling in an almost vacant room, the controller allows a mild temperature rebound, which is consistent with the occupancy-weighted comfort objective defined in Section~\ref{sec:metrics}. 
This behavior is operationally desirable: thermal conditioning is relaxed when the room is scarcely used, thereby avoiding wasteful over-conditioning during low-demand periods.

(3) Once occupants return after 14:00, the controller increases the FCU mode back to Low and Mid, with occasional High-mode actions. 
This recovery behavior is also physically plausible. 
Because Room~6 is embedded in the coupled zone group (Rooms~5--7), the controller must respond not only to the return of internal heat gains from occupants, but also to the thermal influence of adjacent zones. 
Therefore, the temporary increase in actuation after lunch can be interpreted as adaptive recovery.

(4) A similar pattern appears near the end of the workday (18:00--19:00). 
As occupancy decreases again, the FCU mode is stepped down and then remains mostly Off. 
This shows that the controller does not continue aggressive cooling after the operational demand has weakened. 
Instead, it exploits building thermal inertia to realize an energy-saving setback strategy in the final hour of operation. 
At the same time, the temperature trajectory remains within a reasonable band, suggesting that the reduction in actuation does not cause severe thermal deterioration.

This representative case study also illustrates the \emph{smoothness} of the proposed controller in an operational sense. 
Since the control action is discrete and updated every 5 minutes, the FCU signal is naturally stepwise. 
However, the policy still exhibits a clear daily structure: lower actuation during low-occupancy periods and stronger conditioning only when occupancy and thermal load justify it. 
Notably, the controller avoids prolonged High-mode operation and reduces actuation promptly during both lunch break and the end of the workday. 
Overall, the case study complements the aggregate results in Table~\ref{tab:overall_performance} by showing that the performance gains of the hierarchical LLM-RL framework arise from structured, human-aligned, and occupant-centric decision making.

\section{Limitations}
\label{sec:limitations}

\textcolor{black}{
This study also highlights several practical considerations for broader deployment. Although our framework demonstrates the potential in OCC, several limitations remain.
The effectiveness of the LLM component depends on the combination of two sources of knowledge: the general knowledge encoded in the foundation model and the building-specific operational knowledge introduced through supervised fine-tuning on historical building data. 
In this sense, richer and more representative historical operation records can further strengthen feasible-action generation and may provide additional gains for downstream indoor control performance. 
In the future, the availability of open building-operation datasets would be highly valuable for improving domain adaptation, enabling broader benchmarking, and supporting the continued development of LLM-based control methods for buildings. }
In addition, the current study focuses on discrete fan-speed control in a cooling-dominant multi-zone setting, which serves as a practical first step for validating the proposed hierarchical design. 
Extending our framework to continuous control variables, heating seasons, and more diverse building types remains an important direction for future work.

\section{Conclusion}
\label{sec:conclusion}

This paper presents a hierarchical LLM--RL framework for multi-zone HVAC control, in which a fine-tuned LLM generates state-dependent feasible action masks and a masked DQN performs closed-loop optimization within the reduced action space. By integrating historically grounded operational knowledge with reward-driven learning, the proposed framework alleviates the exploration difficulty caused by the large combinatorial action space in multi-zone buildings. In a calibrated 7-zone office-building case study, the method achieved a mean PPD of 7.30\% with daily HVAC energy use of 140.90~kWh, outperforming the vanilla RL baselines in both comfort and energy use and yielding a better overall comfort--energy trade-off than direct LLM control. Further analyses showed that the LLM-guided masks substantially reduce the effective action space, improve training stability, and make LLMs more effective as feasible-action generators than as stand-alone controllers. Overall, the results demonstrate that combining LLM-based semantic guidance with RL-based numerical optimization is a practical and interpretable approach to occupant-centric and energy-efficient HVAC control in multi-zone buildings.

\textcolor{black}{
\section{Acknowledgment}
This work was supported by the National Natural Science Foundation of China under Grant 62192751, partly by the 111 International Collaboration Program of China under Grant BP2018006, and partly by the BNRist Program under Grant No. BNR2019TD01009, and in part by China National Innovation Center of High-Speed Train Project Under Grant CX/KJ-2020-0006.}

\bibliographystyle{unsrt}

\end{document}